\documentclass[11pt]{article}

\usepackage[a4paper,margin=25mm]{geometry}
\usepackage[table]{xcolor}
\usepackage{graphicx}
\usepackage{amsmath}
\usepackage{amssymb}
\usepackage{newtxtext}
\usepackage{newtxmath}
\usepackage{natbib}
\usepackage{authblk}
\usepackage{arydshln}
\usepackage{hyperref}

\graphicspath{{figures/}}
\hypersetup{
    colorlinks = true,
    urlcolor   = blue,
    citecolor  = black,
    linkcolor  = black
}
\setcitestyle{authoryear,round}

\renewcommand{\vec}[1]{\boldsymbol{#1}}
\renewcommand{\d}{\textnormal{d}}
\numberwithin{equation}{section}
\title{Intermittent turbulence in inclined gravity currents}
\author[1]{Lianzheng Cui\thanks{Corresponding author: \href{mailto:lianzheng.cui22@imperial.ac.uk}{lianzheng.cui22@imperial.ac.uk}}}
\author[1]{Graham O. Hughes}
\author[1]{Maarten van Reeuwijk}
\affil[1]{Department of Civil and Environmental Engineering, Imperial College London, London SW7 2AZ, UK}
\date{}

\begin{document}
\maketitle

\begin{abstract}
Inclined gravity currents on shallow slopes can exhibit pronounced turbulence intermittency. Using direct numerical simulations, we investigate this behaviour for a temporal gravity current over a range of initial Reynolds numbers $Re_0$. For $Re_0=2500$ and a slope angle of $0.5^\circ$, the outer layer of the current exhibits large excursions in turbulence intensity and repeated transitions between turbulent and weakly turbulent states. Analysis of the flow energetics reveals that the intermittency is associated with a finite delay between shear production and dissipation of turbulent kinetic energy. During transitional phases, this delay permits a transient amplification of turbulence, which significantly weakens the mean shear by extracting kinetic energy from the mean flow and promoting entrainment-driven layer growth, ultimately leading to relaminarisation. Increasing $Re_0$ reduces the delay and progressively suppresses intermittency, steering the flow towards a more sustained turbulent state. Motivated by these observations, we develop an autonomous delay-differential model based on the coupled evolution of the mean and turbulent kinetic energies. The model reproduces the observed transition from intermittent to sustained turbulence as the delay is reduced and predicts an increased tendency towards intermittency at larger flux Richardson numbers. The results support an interpretation of intermittent turbulence in inclined gravity currents as a delay-induced oscillation arising from the finite adjustment time of turbulence to changes in the mean flow.
\end{abstract}

\section{Introduction}
\label{sec:Intro}
Submarine gravity currents transport sediments, nutrients, and contaminants from shallow coastal regions to the deep ocean, exerting a major influence on seabed evolution and ocean mixing
\citep{simpson1999book,baines2022topographic}. In natural environments, these currents typically evolve at very high Reynolds numbers and propagate over extremely gentle slopes \citep{piper1988slope}, conditions that remain difficult to reproduce either in laboratory experiments or in high-fidelity numerical simulations. Field observations provide an essential complement, but resolving the internal structure and turbulent dynamics remains challenging \citep{selfsharp2019}. 

A longstanding discrepancy exists between laboratory or numerical predictions of mixing and entrainment and the limited available field measurements. A key unresolved question is whether turbulence and the associated entrainment can remain active at large Richardson numbers. Classical entrainment theories, originating from \citet{turner1959} and later parametrised by \citet{turner1986assumption} \citep[see also][]{hallberg2000ET59,jackson2008ET59}, impose a critical Richardson number beyond which entrainment ceases. However, a range of laboratory studies \citep{cenedese2004dense,cenedese2008mixing,cenedese2010new} and field evidence \citep{lauderdale2008field} indicate that turbulent entrainment can persist, even at large Richardson numbers (small slope angles), and exhibit Reynolds-number dependence. \citet{wells2010relationship} attribute the continued mixing at high Richardson number to intermittent burst events, arising from localised regions where low gradient Richardson numbers coincide with high Reynolds numbers, thereby triggering localised turbulent mixing.

This interpretation is consistent with the broader view that, in the presence of (relatively) strong stratification, turbulence may not be continuously sustained but instead occur intermittently \citep{kondo1978burst,rorai2014busrt,deusebio2015intermittency, van2019}. Field observations likewise frequently report intermittent turbulence \citep[see, for example,][]{sun2002intermittent,riley2008intermittent,lozovatsky2010intermittency}. Recent large-scale direct numerical simulations (DNS) on gentle slopes reported by \cite{salinas2020intemigravity} and \cite{zuniga2022newintermi} further suggest that such intermittency in inclined gravity currents is largely confined to the outer layer, while the inner layer remains fully turbulent. These studies suggest the intermittency is associated with an imbalance between buoyancy forcing and total drag, and the flow cycles between so-called accelerating and decelerating regimes. However, the dynamical origin of the outer-layer intermittency and its dependence on Reynolds number remain unclear.

More recently, \cite{cui2025} show that inclined temporal gravity currents at sufficiently high Reynolds numbers evolve towards a regime in which the outer layer is self-similar and the inner layer is quasi-steady, a state referred to as the  \emph{dynamically equilibrated} turbulent regime. Combined with the theory of \cite{van2019} and invoking the weak coupling between the inner and outer layers, they establish theoretical predictions that appear consistent with geophysical gravity currents. However, this equilibrated regime contrasts with the intermittency described above. The objective of the present study is therefore to examine why the currents at small slope angles depart from the dynamically equilibrated regime identified by \cite{cui2025}. 

To address this objective, we conduct DNS of inclined temporal gravity currents for two slope angles ($0.5^\circ$ and $1^\circ$) across a range of Reynolds numbers that captures both intermittent and sustained turbulence. We observe that the intermittent behaviour is associated with a time delay between shear production and dissipation of turbulence, and we develop a time-delay system model that reproduces the observed intermittent dynamics. The remainder of the paper is organised as follows. The numerical set-up and governing equations are presented in \S\ \ref{sec:setup}. The intermittent dynamics and their Reynolds-number dependence are described in \S\ \ref{sec:intermittency}. The time-delay system is developed in \S\ \ref{sec:delaysys}. We finally discuss the results and draw conclusions in \S\S\ \ref{sec:discuss} and \ref{sec:conclusion}.

\section{Case set-up}
\label{sec:setup}

We consider an inclined temporal gravity current, in which a negatively buoyant layer of height $h_0$, uniform initial buoyancy $b_0$ and velocity $u_0$, descends an infinitely long slope of constant angle $\alpha$. In an inclined coordinate system, with $x$ and $z$ denoting the along-slope and slope-normal directions, respectively, the governing equations under the Boussinesq approximation are
\begin{align}
  \label{eq:momentum equation}
  \frac{\partial \bf{u}}{\partial t}+{\bf u} \cdot \nabla {\bf u}&=-\nabla p+v\nabla^2{\bf u}
  +b\mathbf{e}_g, \\
  \label{eq:continuity equation}
  \nabla\cdot{\bf u}  &=0,
\end{align}
where $\mathbf{u}=(u,v,w)$ is the velocity vector, $p$ is the kinematic pressure, $\nu$ is the kinematic viscosity, $b=(\rho_a-\rho)g/\rho_a$ denotes the buoyancy and $\rho_a$ is the ambient reference density. The unit vector $\mathbf{e}_g=(-\sin\alpha,0,\cos\alpha)$ is aligned with the gravitational acceleration. The transport of buoyancy is governed by
\begin{equation}
    \label{eq:buoyancy transport}
  \frac{\partial b}{\partial t}+{\bf u} \cdot \nabla b =\kappa\nabla^2 b,
\end{equation}
where $\kappa$ is the kinematic diffusivity.

To simulate this unbounded problem in a finite domain, we impose periodic boundary conditions in the streamwise ($x$) and spanwise ($y$) directions of the domain, with free-slip and no-slip wall conditions at the upper and lower boundaries, respectively. This configuration yields statistical homogeneity in the horizontal plane so that  horizontally averaged quantities depend only on the slope-normal coordinate $z$ and time $t$. The temporal evolution of the flow corresponds closely to the body of a spatially evolving current in terms of entrainment-driven growth, flow structure and self-similarity \citep{van2019, cui2025}. 

Within this temporal framework, horizontally averaging the governing equations for the streamwise velocity $u$ and buoyancy $b$ yields the Reynolds-averaged equations
\begin{align}
 \frac{\partial\overline{u}}{\partial t} +\frac{\partial\overline{w'u'}}{\partial z}&=v\frac{\partial^2\overline{u}}{\partial z^2}-\overline{b}\sin\alpha, 
 \label{eq:averaged momentum} \\
 \frac{\partial\overline{b}}{\partial t} + \frac{\partial\overline{w'b'}}{\partial z}&=\kappa\frac{\partial^2\overline{b}}{\partial z^2},
  \label{eq:averaged buoyancy} 
\end{align}
where $\overline{\chi}=\iint \chi~\d x \d y/(L_xL_y)$ represents the spatial averaging operator for the quantity $\chi$, $L_x$ and $L_y$ are the dimensions of the domain in the $x$ and $y$ directions, respectively, and a prime represents the departure from the corresponding planar average, i.e. $\chi' = \chi - \overline{\chi}$.

Since $\overline w = 0$ due to continuity and $\overline v = 0$ due to the absence of forcing in that direction, multiplying \eqref{eq:averaged momentum} by $\overline{u}$ yields the mean kinetic energy (MKE) equation. Taking the dot product of the velocity fluctuations $\boldsymbol{u'}$ with \eqref{eq:momentum equation} and averaging over the homogeneous $x$ and $y$ directions gives the turbulent kinetic energy (TKE) equation. The balance equations for MKE $\overline{e}=\overline{u}^2/2$ and TKE $e=\overline{u_i'^2}/2$ may be written as
\begin{align}
\label{eq:MKE}
  \frac{\partial \overline{e}}{\partial t}&= \underbrace{\overline{w'u'}\frac{\partial\overline{u}}{\partial z}}_{-P_S}\underbrace{ -\overline{u}\overline{b}\sin\alpha}_{\overline{P}_B}-\underbrace{\nu\left(\frac{\partial\overline{u}}{\partial z}\right)^2}_{\overline{\varepsilon}}+\underbrace{\frac{\partial}{\partial z}\left(-\overline{u}\overline{w'u'}+\nu\frac{\partial\overline{e}}{\partial z}\right)}_{T_M},\\
\label{eq:TKE}
  \frac{\partial e}{\partial t}&=\underbrace{- \overline{w'u'} \frac{\partial \overline u}{\partial z}}_{P_S}\underbrace{-\overline{u'b'} \sin \alpha + \overline{w'b'} \cos\alpha }_{P_B}-\underbrace{\nu\overline{\left(\frac{\partial u_i'}{\partial x_j}\right)^2}}_{\varepsilon}+\underbrace{\frac{\partial}{\partial z}\left(-\frac{\overline{w'u_i'u_i'}}{2}-\frac{\overline{w'p'}}{\rho_a}+\nu\frac{\partial e}{\partial z}\right)}_{T_T}.
\end{align}
Here, $P_S$ denotes shear production of TKE, which converts energy from MKE to TKE. $\overline{P}_B$ and $P_B$ represent buoyancy production of MKE and TKE, respectively, $\overline{\varepsilon}$ and $\varepsilon$ are the respective dissipation rates of MKE and TKE, and $T_M$ and $T_T$ denote the respective transport terms. To characterise the outer-layer dynamics, we define the volume flux $Q$, momentum flux $M$,  integral TKE $K$  and integral buoyancy forcing $B$ of the outer layer as
\begin{equation*}
    Q = \int_{z_{um}}^\infty \overline{u} \d z,\quad
    M = \int_{z_{um}}^\infty \overline{u}^2 \d z,\quad 
    K=\int_{z_{um}}^\infty e \d z,\quad
    B = \int_{z_{um}}^\infty -\overline{b}\sin\alpha\d z, 
    \refstepcounter{equation}
    \eqno{(\theequation{a-d})}
    \label{eq:IntQuan}
\end{equation*}
where $z_{um}$ is the vertical coordinate at which $\overline{u}$ attains its maximum value. The characteristic outer-layer length scale $h$, velocity $u_T$, buoyancy $b_T$, MKE $\overline{e}_T$ and TKE $e_T$ are then defined similarly to  \cite{van2019}:
\begin{equation*}\label{eq:hutet}
    h=\frac{Q^2}{M},\quad u_T=\frac{Q}{h},\quad b_T=-\frac{B}{h\sin\alpha},\quad \overline{e}_T=\frac{1}{2}\frac{M}{h}=\frac{1}{2}u_T^2,\quad e_T=\frac{K}{h}.
    \refstepcounter{equation}
    \eqno{(\theequation{a-e})}
\end{equation*}
The outer-layer dimensionless parameters, including the Reynolds number $Re$, bulk Richardson number $Ri$, buoyancy Reynolds number $Re_b$, and Taylor-scale Reynolds number $Re_\lambda$ are defined as
\begin{equation*}\label{eq:dimless}
   Ri=-\frac{b_Th\cos\alpha}{u_T^2},\quad Re = \frac{u_Th}{\nu},\quad {Re_b}=\frac{\varepsilon_T}{\nu N^2},\quad Re_\lambda = \sqrt{\frac{20}{3}\frac{e_T^2}{\varepsilon_T\nu}},
    \refstepcounter{equation}
    \eqno{(\theequation{a-d})}
\end{equation*}
where $\varepsilon_{T} = h^{-1}\int_{z_{um}}^\infty \varepsilon \d z$ denotes the characteristic dissipation rate of TKE and ${N} = \sqrt{-9b_T\cos\alpha/(8h)}$ is an outer-layer characteristic buoyancy frequency, derived under the assumption of a linear buoyancy profile (see \cite{van2019} for details). 

\begin{table}
  \begin{center}
\def~{\hphantom{0}}
  \begin{tabular}{lccccccccc}
      $Sim.$  & $\alpha$    &$Ri_0$& $Re_0$&$N_x\cdot N_y\cdot N_z$&$t_{sim}/t^*$&$\hat{\tau}/t^*$& $Re_b$& $Re_\lambda$ \\[3pt]
       S0.5R25 & $0.5^\circ$ &2.22&2500&~$1024^3$&305&1.32&47.11&44.87\\
       S0.5R37 & $0.5^\circ$ &2.22&3700 &~$1536^3$&240&1.22&72.16&52.37\\
       S0.5R50 & $0.5^\circ$ &2.22&5000 &~$2048^3$&240&1.19&99.21&60.73\\
       S1R25 ~&~$1^\circ$~  &~1.11~& 2500 &~$1024^3$&240&1.30&54.15&42.13 \\
  \end{tabular}
  \caption{Simulation details. The Prandtl number $Pr = \nu/\kappa$ is 1 and the computational domain has a size of $(20h_0)^3$ for all cases.}
  \label{tab:sim}
  \end{center}
\end{table}

The principal simulation parameters are listed in table \ref{tab:sim}. The parameters $Ri_0=-b_0h_0\cos\alpha/u_0^2$ and $Re_0 = u_0h_0/\nu $ are the initial Richardson and Reynolds numbers, respectively. We vary $Re_0$ at a fixed slope angle $\alpha = 0.5^\circ$ ($Ri_0=2.22$) to investigate the Reynolds-number dependence of the  dynamics governing intermittency. Case S1R25 ($\alpha = 1^\circ$, $Re_0 = 2500$) serves as a reference case without sustained intermittency (aside from an initial adjustment), consistent with case 1N of \citet{cui2025}. The grid resolution $N_x \cdot N_y \cdot N_z$ is chosen to ensure that $\Delta x / L_k \lesssim 1.5$ in all cases, where $\Delta x$ is the grid spacing (uniform in the three directions) and $L_k = (\nu^3/\varepsilon_T)^{1/4}$ is the characteristic Kolmogorov length scale. The simulations are run for a duration $t_{sim}$ to examine the evolution of turbulence. The characteristic timescale is defined as $t^* = h_0/\sqrt{B_0}$, where $B_0 = \int-\overline{b}\sin\alpha \d z$ is the integral buoyancy forcing (remaining constant due to buoyancy conservation). The delay timescale $\hat{\tau}$ characterises the lagged response of TKE dissipation to changes in shear production, as obtained from the cross-correlation analysis in  \S\ \ref{sec:intermittency}. The observed Reynolds numbers $Re_b$ and $Re_\lambda$ in table \ref{tab:sim} are averaged over the transition period (e.g. $t/t^*\in[140,160]$ for case S0.5R25; see figure \ref{fig:TKEevolution}$(b)$ for the corresponding development of turbulence during this interval). The Prandtl number $Pr = \nu/\kappa$ is fixed at 1 and the computational domain is $(20h_0)^3$ for all cases. 

All simulations are performed with the in-house DNS code SPARKLE, which has been extensively validated for gravity-current flows and is described in detail by \cite{craske2015jet1}. Further details of the simulations and the comparison between temporal and spatial formulations are provided in \citet{van2019, cui2025}.

\section{Turbulence intermittency}\label{sec:intermittency}
\subsection{Intermittency at low slope angles}\label{sec:evolution}
\begin{figure}
  \centering
   \includegraphics[scale =0.26, trim=0.25cm 0.25cm 0.25cm 0cm, clip] {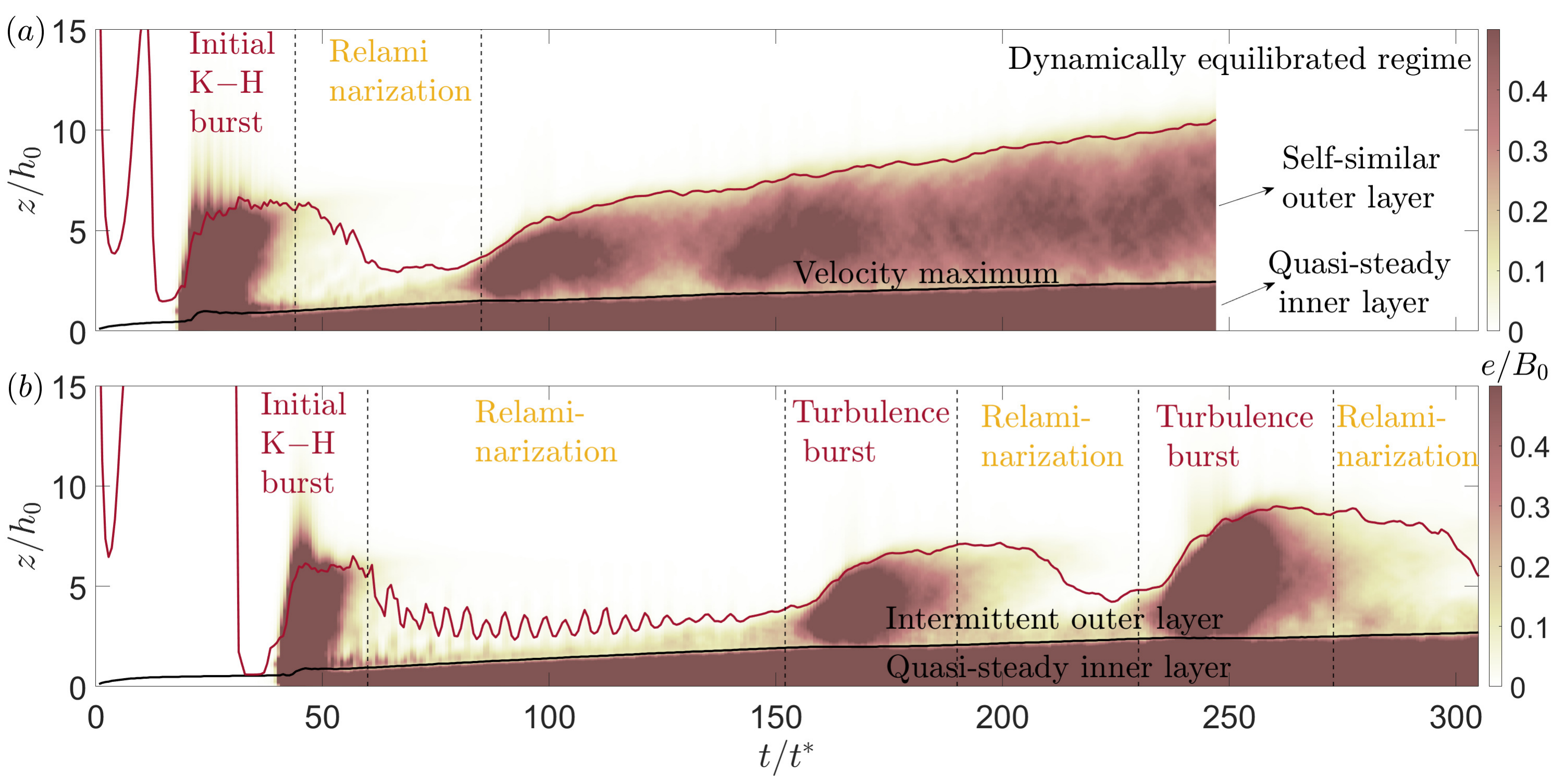}
  \caption{Temporal variation of the dimensionless turbulent kinetic energy, $e/B_0$, for $(a)$ case S1R25 ($\alpha = 1^\circ$) and $(b)$ case S0.5R25 ($\alpha = 0.5^\circ$). The black solid lines indicate the level of $z_{um}$. The red solid lines show the characteristic thickness of the turbulent region in the outer layer, $h_e$, measured from $z_{um}$. The vertical lines delineate the laminar and turbulent phases in the outer layer. }
 \label{fig:TKEevolution}
\end{figure}

Figure \ref{fig:TKEevolution}$(a)$ demonstrates the evolution of turbulence for case S1R25 ($\alpha = 1^\circ$), quantified by the TKE $e$ (normalised by $B_0$). The black solid line marks the level of the velocity maximum $z_{um}$, where the shear production of TKE vanishes and the inner and outer layers are clearly separated. Following an initial turbulent burst (associated with the Kelvin--Helmholtz (K--H) instability) and a subsequent relaminarisation, the outer layer evolves into a turbulent regime with relatively constant TKE levels and entrainment rates. The red solid line indicates the height $h_e + z_{um}$, where $h_e = K^2/\int_{z_{um}}^\infty e^2 \d z$ may be interpreted as a characteristic measure of the thickness of the turbulent region in the outer layer. This height is seen to increase approximately linearly with time due to turbulent entrainment. Meanwhile, turbulence in the inner layer is established immediately after the initial burst, and the layer depth increases only gradually with time. These dynamics are consistent with the dynamically equilibrated regime reported by \citet{cui2025}, in which the outer layer evolves in an approximately self-similar manner while the inner layer approaches a quasi-steady state.

In contrast to case S1R25, halving the slope (doubling $Ri_0$) while keeping the initial Reynolds number $Re_0$ fixed profoundly alters the behaviour of the outer layer. Figure \ref{fig:TKEevolution}$(b)$ shows the corresponding evolution for case S0.5R25 ($\alpha=0.5^\circ$). Here, following the initial burst and relaminarisation, the outer layer fails to reach a persistent self-similar regime. Instead, it exhibits pronounced intermittency, transitioning between turbulent bursts and relaminarisation phases (delineated by the vertical dashed lines in figure \ref{fig:TKEevolution}). The thickness of the turbulent region $h_e + z_{um}$ expands significantly during burst phases, then contracts during relaminarisation periods. Remarkably, despite this dramatic change in outer layer dynamics, the inner layer follows a trajectory comparable to case S1R25, reaching a quasi-steady state soon after the initial burst and maintaining turbulence throughout the subsequent evolution. This demonstrates the robustness of the near-wall turbulent structures and confirms the weak coupling between the two layers even under these modified conditions.
\begin{figure}
  \centering
   \includegraphics[scale =0.35, trim=4cm 2cm 8cm 0cm, clip] {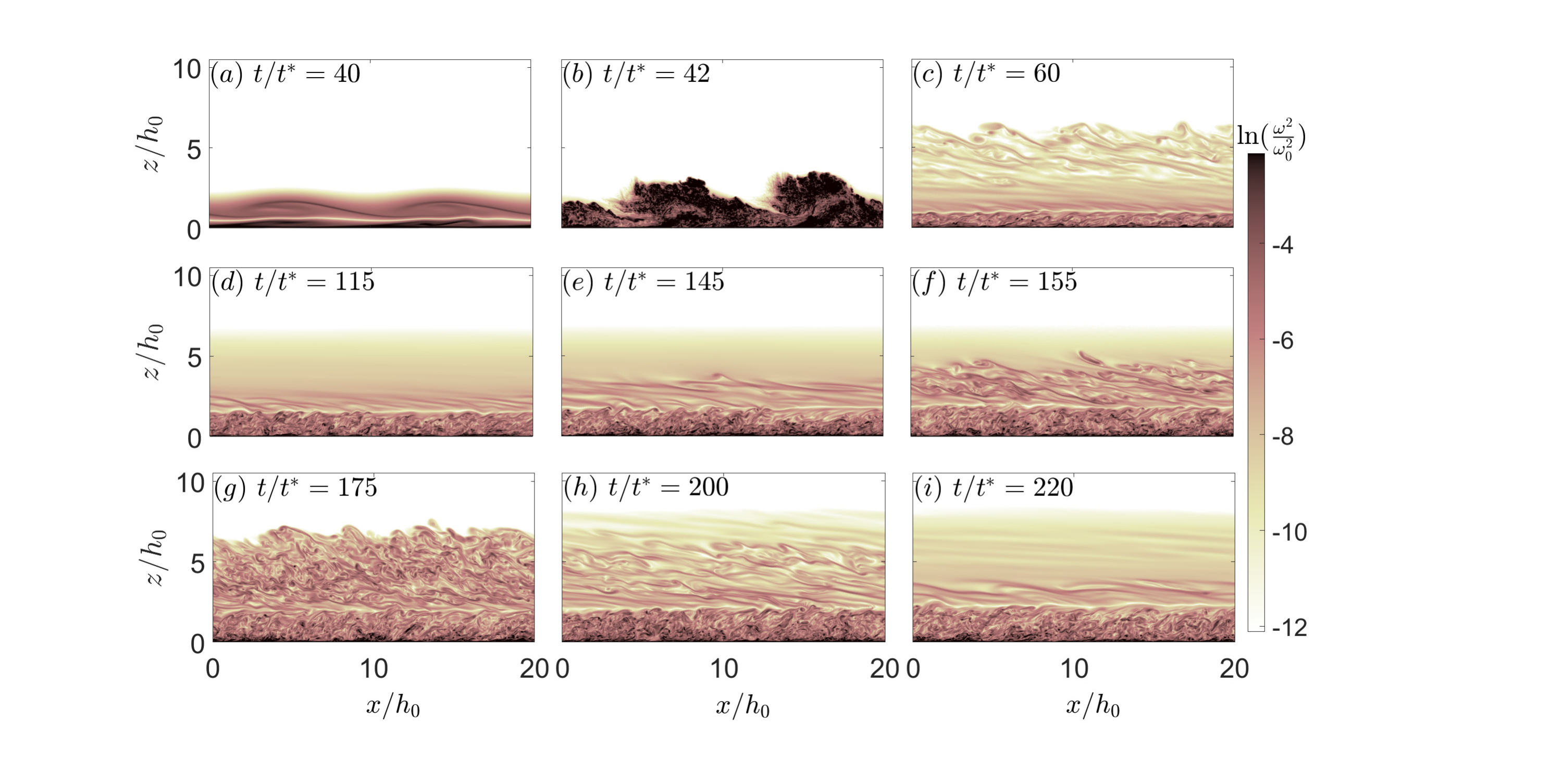}
  \caption{Instantaneous enstrophy for case S0.5R25 shown on a logarithmic scale $\ln(\omega^2/\omega_0^2)$ for times ranging from $(a)\ t/t^* = 40$ to $(i)\ t/t^* = 220$. The corresponding times are indicated in each panel.}
 \label{fig:enstrophy}
\end{figure}

The turbulence structure of the intermittent case (S0.5R25) is further elucidated through instantaneous fields of the normalised enstrophy density,  shown in terms of the quantity $\ln(\omega^2/\omega_0^2)$ at selected times in figure \ref{fig:enstrophy}. Here, $\omega^2=\omega_i\omega_i$ denotes the local enstrophy density, with $\omega_i$ the components of the vorticity vector, and $\omega_0^2=u_0^3/(h_0\nu)$ is a reference value. The initial (K--H) instability sequence is captured in panels $(a-c)$, where intense vortical structures are seen to develop rapidly in the current. Panel $(d)$ shows the subsequent relaminarisation phase, with turbulence confined primarily to the inner layer. The regeneration of outer layer turbulence is evident in the progression from panels $(d)$ to $(g)$, where eddies reoccupy the outer layer, culminating in panel $(g)$ where the outer layer becomes fully turbulent again. A second suppression follows this turbulent regime, restoring a quiescent outer state as shown in panels $(h)$ and $(i)$.  Throughout, the inner shear layer remains turbulent and slowly deepening, in line with the quasi-steady state and the weak inner–outer coupling.

\begin{figure}
  \centering
    \includegraphics[scale =0.28, trim=0.3cm 1.5cm 1cm 0cm, clip] {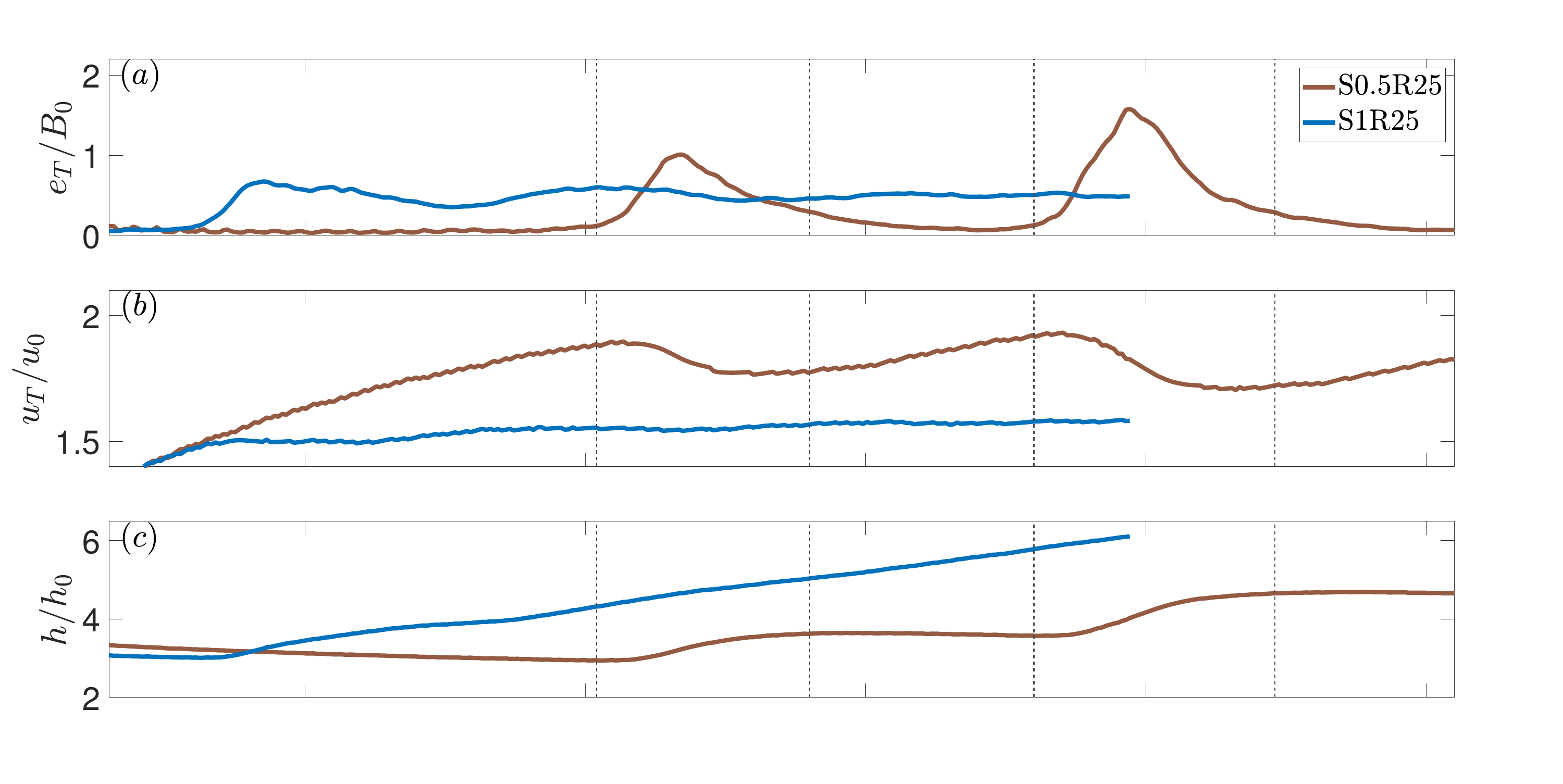}
    \includegraphics[scale =0.28, trim=0.3cm 15cm 1cm 1.5cm, clip] {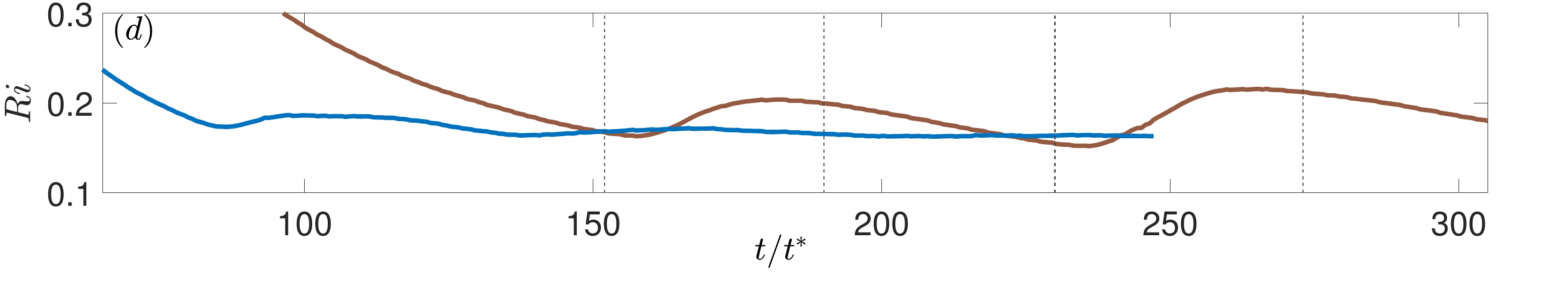}
  \caption{Temporal variation of normalised outer-layer characteristic quantities: $(a)$ $e_T/B_0$, $(b)$ $u_T/u_0$, $(c)$ $h/h_0$ and $(d)$ $Ri$, for cases S1R25 and S0.5R25. The vertical lines delineate the laminar and turbulent phases.} 
 \label{fig:Characteristic values}
\end{figure}

Since pronounced intermittency is confined to the outer layer, the following analysis focuses on its evolution. Figure \ref{fig:Characteristic values} shows the temporal variation of the characteristic outer-layer quantities $e_T$, $u_T$, $h$ (normalised by their respective initial values), and $Ri$ for cases S1R25 (blue) and S0.5R25 (brown). The vertical black lines mark the transitions from turbulent bursts to relaminarisation. For case S1R25 ($\alpha = 1^\circ$), $e_T$, $u_T$ and $Ri$ remain approximately constant once the flow reaches the self-similar regime, while $h$ increases nearly linearly in time, consistent with sustained turbulence.

In contrast to case S1R25, case S0.5R25 ($\alpha=0.5^\circ$) displays clear signatures of intermittency. The quantities $e_{T}$, $u_{T}$ and $Ri$ undergo pronounced excursions, directly reflecting the alternating laminar and turbulent phases observed in figure \ref{fig:TKEevolution}($b$). During laminar phases ($e_T \approx 0$), $h$ remains nearly constant, while $u_T$ increases approximately linearly in time under constant buoyancy forcing, as suggested by the reduced form of \eqref{eq:averaged momentum}: $\partial \overline{u}/\partial t = -\overline{b}\sin\alpha$ in the absence of turbulence. The resulting increase in shear leads to a progressive reduction in $Ri$ until a subsequent turbulent burst is triggered. Compared with S1R25, case S0.5R25 requires a longer acceleration phase prior to transition, since a stronger mean shear ($u_T/h$) is needed to overcome the stronger stratification at the smaller slope angle. During turbulent phases, $e_T$ peaks and $h$ increases correspondingly due to turbulent entrainment, whereas $u_T$ decreases as Reynolds stresses begin to outweigh buoyancy forcing (see \eqref{eq:averaged momentum}). Notably, $Ri$ decreases to approximately $0.15$ near the delineated transition-to-turbulence times for both burst events. The threshold is close to the critical value reported by \citet{van2019}, above which turbulence may not be sustained.

\begin{figure}
  \centering
    \includegraphics[scale =0.28, trim=3.5cm 2.5cm 2cm 0cm, clip] {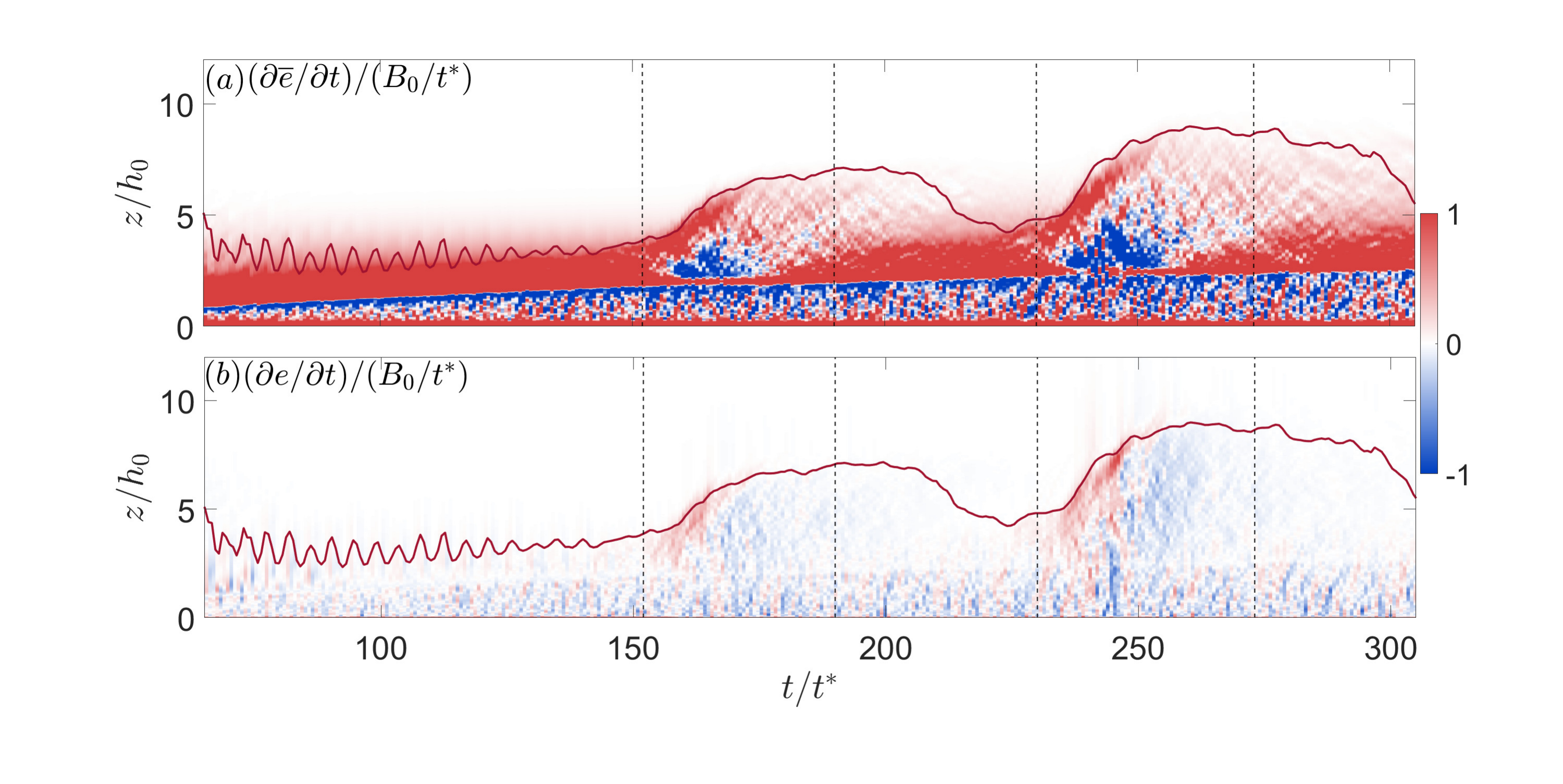}
  \caption{Time derivatives of $(a)$ the mean kinetic energy $\partial \overline{e}/\partial t$ and $(b)$ the turbulent kinetic energy $\partial e/\partial t$, both normalised by $B_0/t^*$. The red solid lines show the characteristic thickness of the turbulent region in the outer layer, $h_e$, measured from $z_{um}$. The vertical lines delineate the laminar and turbulent phases.}
 \label{fig:MTKE}
\end{figure}

\subsection{Non-equilibrium energy budgets}\label{sec:energetics}
To elucidate the intermittent dynamics at low slope angles, we examine the flow energetics through the budgets of mean kinetic energy (MKE) and turbulent kinetic energy (TKE). Figure \ref{fig:MTKE} illustrates normalised $\partial \overline{e}/\partial t$ and $\partial e/\partial t$ for case S0.5R25. Panel $(a)$ shows that MKE increases during laminar phases with negligible vertical redistribution. In contrast, during turbulent phases, the MKE field is seen to undergo a pronounced reorganization: turbulent entrainment drives the upward advance of MKE and leads to the reduction of MKE in the lower portion of the outer layer. Following the decay of turbulence, the MKE field returns to a buoyancy-driven growth regime. The evolution of TKE (figure~\ref{fig:MTKE}$(b)$) exhibits  a clear growth–decay sequence throughout the outer layer during the turbulent phase: turbulence initially intensifies as shear production transfers energy from the mean flow, and subsequently decays after the turbulent region in the outer layer has undergone a rapid expansion.

To better quantify the individual contributions within the MKE and TKE budgets, we integrate \eqref{eq:MKE} and \eqref{eq:TKE} over the outer layer $(z_{um},\infty)$ and note $\d z_{um}/\d t \approx 0$ (consistent with the weak growth of the inner layer shown in figure \ref{fig:TKEevolution}), yielding
\begin{align}
\label{eq:debtdt}
  \frac{\d \overline{e}_T}{\d t}&= -\frac{\d h}{\d t}\frac{\overline{e}_T}{h}-{{P}_{ST}}+\overline{{P}}_{BT},\\
\label{eq:detdt}
  \frac{\d e_T}{\d t}&= -\frac{\d h}{\d t}\frac{e_T}{h}+{{P}_{ST}}+{P}_{BT}-{\varepsilon}_{T},
\end{align}
where $P_{ST}=h^{-1}\int_{z_{um}}^\infty P_s \d z$ is the characteristic shear production, $\overline{P}_{BT} = h^{-1}\int_{z_{um}}^\infty \overline{P}_B \d z$ and $P_{BT}= h^{-1}\int_{z_{um}}^\infty P_B \d z$ denote the characteristic buoyancy production of MKE and TKE, respectively. The dissipation of MKE has been omitted from  \eqref{eq:debtdt}, as $\overline{\varepsilon}$ is negligible in the outer layer. The transport terms $T_M$ and $T_T$ are also absent, since they primarily redistribute energy within the inner and outer layers separately, consistent with the weak coupling between the two layers, as observed in  \citet{cui2025}.
\begin{figure}
  \centering
   \includegraphics[scale =0.28, trim=0.5cm 14.5cm 1cm 0cm, clip] {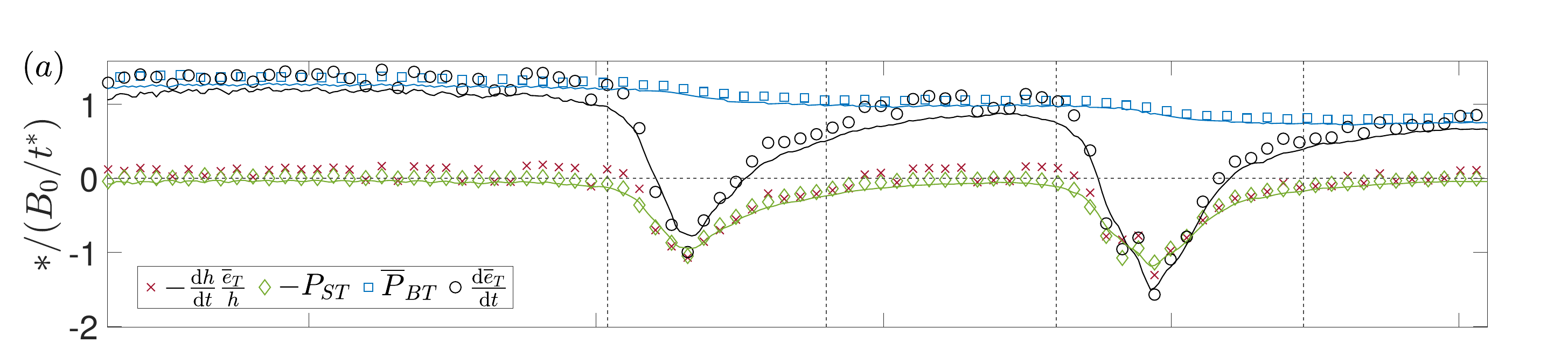}
     \includegraphics[scale =0.28, trim=0.5cm 0.5cm 1cm 13.5cm, clip] {IntBudgets_new-eps-converted-to.pdf}
  \caption{Time variation of the normalised budget terms for the outer-layer characteristic $(a)$ mean kinetic energy $\overline{e}_T$ and $(b)$ turbulent kinetic energy $e_T$. The solid lines denote the model predictions developed in \S\S\ \ref{sec:energetics} and \ref{sec:sys}, with each prediction shown in the same colour as the corresponding budget term.}
 \label{fig:ebartet}
\end{figure}

Figure \ref{fig:ebartet} shows the temporal variation of normalised $\d \overline{e}_T/\d t$ and $\d e_T/\d t$, with contributions from the similarly normalised terms in \eqref{eq:debtdt} and \eqref{eq:detdt}. Panel $(a)$ shows $\d \overline{e}_T/\d t$ (black circles) and the contributing terms. The characteristic buoyancy production $\overline{P}_{BT}$ (blue squares) dominates the balance and closely matches $\d \overline{e}_T/\d t$ in the laminar phase. The magnitude of $\overline{P}_{BT}$ decreases gradually in time, with the most pronounced reductions occurring in a step-like manner following turbulent events. Assuming that $\overline{b}$ and $\overline{u}$ share a similar vertical structure in the outer layer, $\overline{P}_{BT}$ can be modelled as
\begin{equation}
    \overline{P}_{BT} \approx {B u_T}/{h}.
    \label{eq:Pbarbt}
\end{equation}
A detailed derivation is provided in Appendix \ref{sec:barPBT}. This estimate is shown by the blue solid line in figure \ref{fig:ebartet}$(a)$ and is in good agreement with the DNS data throughout the evolution, including during the turbulent burst phases. This suggests that, to a good approximation, the momentum and buoyancy fields are rapidly redistributed in a similar manner in the outer layer during these bursts.

During the turbulent phases, $\d\overline{e}_T/\d t$ becomes negative, driven by the characteristic shear production $P_{ST}$ (green diamonds) and the geometric dilution term $-(\d h/\d t)(\overline{e}_T/h)$ (red crosses). The former represents the conversion of $\overline{e}_T$ to $e_T$, while the latter accounts for the dilution of $\overline{e}_T$ by the outer layer expansion. Remarkably, these two terms make approximately equal contributions to the evolution of $\overline{e}_T$ in \eqref{eq:debtdt}. This relation is obtained by substituting \eqref{eq:averaged momentum} into (\ref{eq:IntQuan}$a$), using (\ref{eq:IntQuan}$d$), and noting that $\d z_{um}/\d t \approx 0$, which yields
\begin{equation}
    \frac{\d Q}{\d t}=u_{T}\frac{\d h}{\d t}+h\frac{\d u_{T}}{\d t}\approx B.
    \label{eq:dQdt}
\end{equation}
Rearranging \eqref{eq:dQdt} for $\d u_T/\d t$, multiplying by $u_T$, and substituting both the resulting relation and \eqref{eq:Pbarbt} into \eqref{eq:debtdt} (noting that $\d  \overline{e}_T/\d  t = u_T\d u_T/\d t$) gives
\begin{equation}
    \label{eq:dhdt}
    \frac{\d h}{\d t}\approx\frac{2P_{ST}h}{u_T^2}\qquad\Rightarrow\qquad  \frac{\overline{e}_T}{h}\frac{\d h}{\d t}\approx P_{ST},
\end{equation}
confirming that the two contributions are approximately equal. Turbulence thus weakens the mean shear through two equally important pathways: by extracting energy from the mean flow and by diluting that energy through entrainment-induced thickening of the outer layer.

Figure \ref{fig:ebartet}$(b)$ shows the evolution of $\d e_T/\d t$ (black circles) together with its individual budget components. The leading-order balance is maintained between $P_{ST}$ (green diamonds) and the characteristic dissipation $\varepsilon_T$ (yellow circles). The buoyancy-production term associated with turbulent fluctuations, $P_{BT}$ (blue circles), exhibits small negative values, indicating a weak return of TKE to potential energy. The geometric-dilution term $-(\d h/\d t)(e_T/h)$ (red circles) plays only a secondary role. The unsteady term $\d e_T/\d t$ exhibits approximate antisymmetry during the turbulent phase, reflecting a minor mismatch between $P_{ST}$ and $\varepsilon_T$. This is presumably because a finite time lag exists between the energy input and the dissipation, representing the characteristic time required for dissipation to adjust to changes in the mean flow. This is discussed further below. Consistent with this picture, $\varepsilon_T$ is observed to lag slightly behind $P_{ST}$ in figure \ref{fig:ebartet}$(b)$, allowing production to exceed dissipation temporarily (supporting turbulence growth) and later to fall below it (leading to turbulence decay).

\subsection{Reynolds number dependence}\label{sec:Redepdent}
\begin{figure}
  \centering
   \includegraphics[scale =0.28, trim=2cm 2cm 2cm 1.5cm, clip] {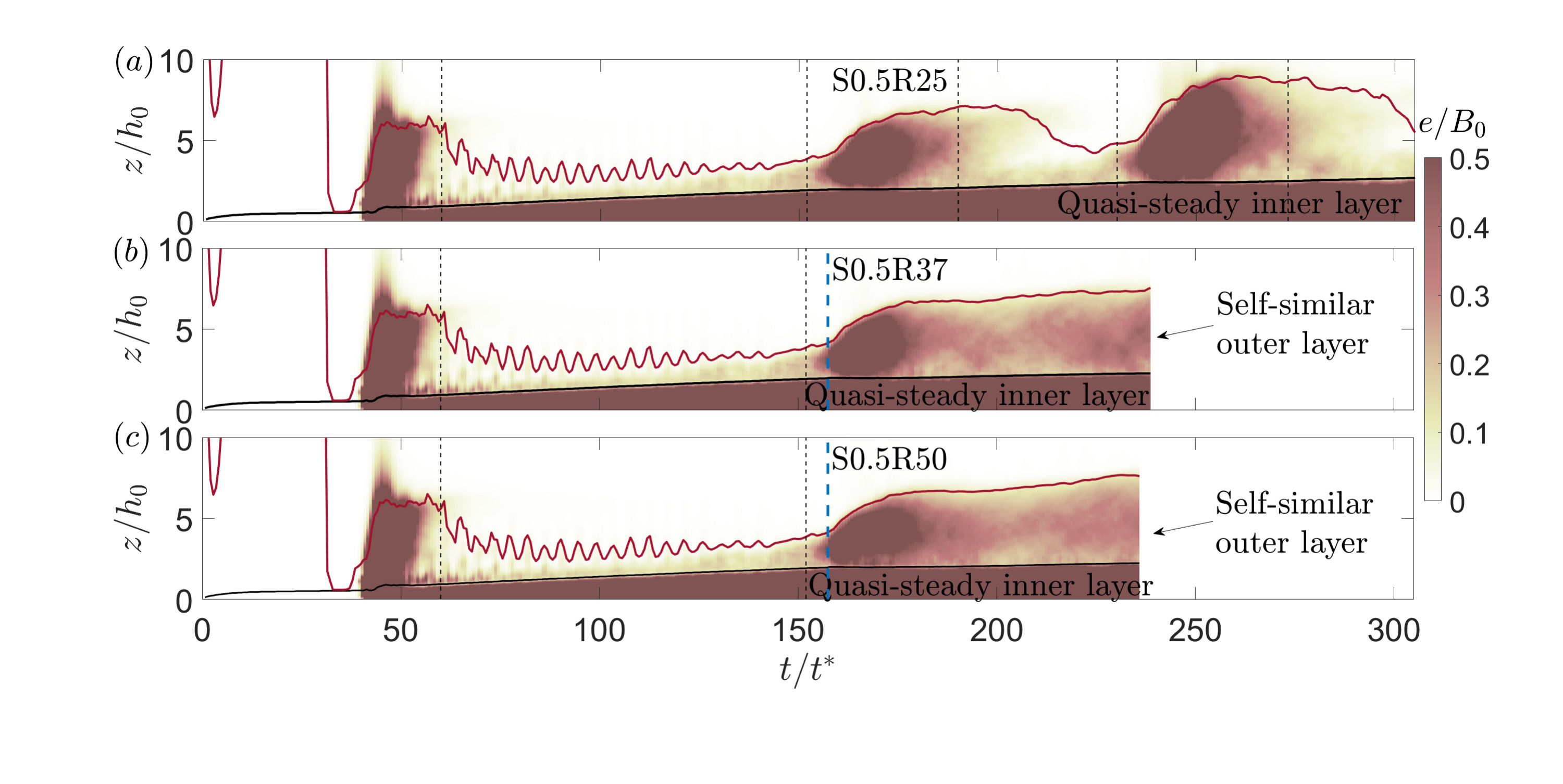} 
  \caption{Temporal variation of the normalised turbulent kinetic energy, $e/B_0$, for cases $(a)$ S0.5R25 ($Re_0=2500$), $(b)$ S0.5R37 ($Re_0=3700$) and $(c)$ S0.5R50 ($Re_0=5000$). The red solid lines show the level of $h_e + z_{um}$. The blue vertical lines mark the restart points for cases S0.5R37 and S0.5R50. }
 \label{fig:DifReEvo_2D}
\end{figure}
\begin{figure}
  \centering
   \includegraphics[scale =0.28, trim=0cm 0cm 1cm 0cm, clip] {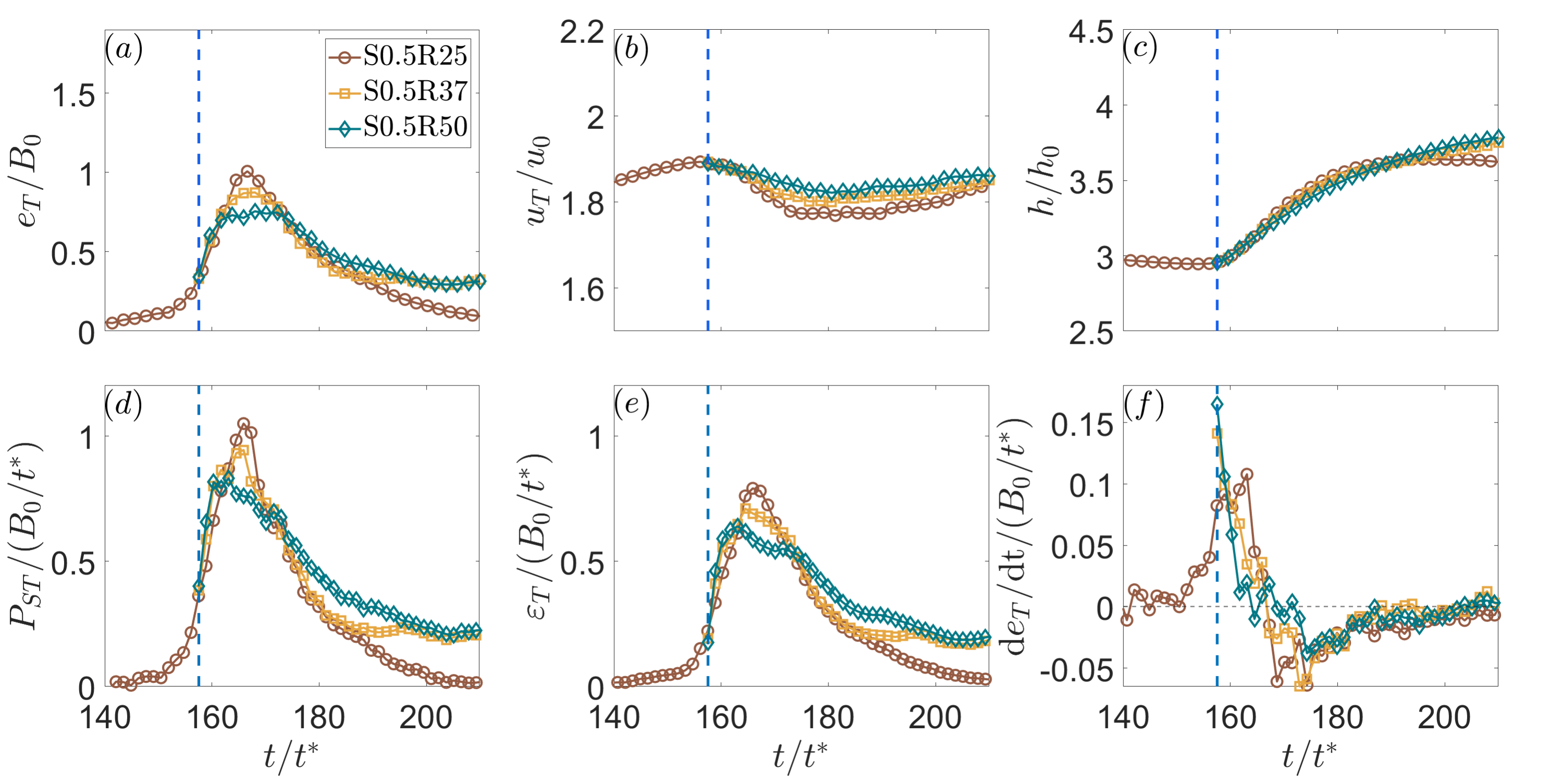}
  \caption{Temporal variation of normalised outer-layer characteristic quantities: $(a)$ $e_T$, $(b)$ $u_T$, $(c)$ $h$, $(d)$ $P_{ST}$, $(e)$ $\varepsilon_T$ and $(f)$ $\d e_T/ \d t$, for cases S0.5R25, S0.5R37 and S0.5R50. The blue vertical line marks the restart point for cases S0.5R37 and S0.5R50.}
 \label{fig:DifReEvo_Chrac}
\end{figure}

In this section, we investigate the dependence of the dynamics governing intermittency on the  Reynolds number. We systematically vary $Re_0$ at $\alpha=0.5^\circ$ by adjusting the kinematic viscosity $\nu$. Specifically, for the higher-$Re$ cases S0.5R37 and S0.5R50, the simulations are restarted from a common flow field extracted from case S0.5R25 (at $t/t^*=157$). At this restart point, $\nu$ is reduced while all other physical parameters are kept identical, thereby increasing $Re_0$ without altering the large-scale flow properties. To ensure adequate resolution of the smallest turbulence scales, the computational grid is refined accordingly in cases S0.5R37 and S0.5R50 (see table \ref{tab:sim} for details). The flow field from the coarser grid is interpolated onto the refined grid using a standard trilinear interpolation. The subsequent adjustment to the higher Reynolds number is expected to occur over approximately $t^*$, comparable to the mean production--dissipation lag obtained from the cross-correlation analysis presented below. This rapid adjustment is also evident in figure \ref{fig:DifReEvo_Chrac}$(e)$, where, despite the reduced viscosity, the dissipation at the higher $Re_0$ quickly exceeds that at the lower $Re_0$ following the restart.

Figure \ref{fig:DifReEvo_2D} shows the evolution of TKE for the three values of  $Re_0$ at $\alpha = 0.5^\circ$, where the blue vertical lines mark the restart points (indicated by the blue vertical dashed lines in panels $(b)$ and $(c)$) at which $\nu$ is reduced and cases S0.5R37 and S0.5R50 are initialised. Panel $(a)$ presents the intermittent case S0.5R25 for reference.  As $Re_0$ increases, it is apparent that turbulence becomes more sustained after the burst event. Although moderate fluctuations persist for higher $Re_0$, the thickness of the turbulent region (red solid lines) increases gradually rather than undergoing the sharp contraction evident in panel $(a)$. The outer layer maintains continuous turbulent activity and evolves towards a self-similar regime, resembling that observed in case S1R25 (see figure \ref{fig:TKEevolution}$(a)$). The results indicate that increasing the Reynolds number enhances the robustness of turbulence, as discussed further below.

Figure \ref{fig:DifReEvo_Chrac} presents the evolution of the characteristic quantities for the cases at $\alpha = 0.5^\circ$. The blue vertical lines again mark the time at which cases S0.5R37 and S0.5R50 are initiated. Panel $(a)$ shows that, counter-intuitively, the lowest-$Re_0$ case (S0.5R25) exhibits the largest TKE peak during turbulent bursts, whereas the peaks in S0.5R37 and S0.5R50 decrease with increasing $Re_0$. A similar trend appears in panel $(b)$: the low-$Re_0$ case undergoes the largest (but relatively modest) reduction in $u_T$, indicating a stronger proportionate decrease in MKE. The normalised layer thickness in panel $(c)$ remains similar across the range of $Re_0$, although the thickness in case S0.5R25 is slightly greater than in the other two cases before turbulence decay sets in.

To examine why bursts are more pronounced at lower $Re_0$, figure \ref{fig:DifReEvo_Chrac}$(d-f)$ compares the normalised $P_{ST}$, $\varepsilon_T$, and $\d e_T / \d t$, respectively. Both $P_{ST}$ and $\varepsilon_T$ attain progressively smaller peaks with increasing $Re_0$, consistent with the behaviour of $e_T$. However, during the initial turbulence-growth stage, $P_{ST}$ and particularly $\varepsilon_T$ grow more rapidly in the higher-$Re_0$ cases. The unsteady term $\d e_T / \d t$ likewise shows a larger initial magnitude at higher $Re_0$, but decays more rapidly. This behaviour suggests that, at lower $Re_0$, $\varepsilon_T$ may respond more slowly to changes in $P_{ST}$, allowing a temporary imbalance between production and dissipation and hence a transient accumulation of TKE. The associated increase in $e_T$ may then weaken the mean shear to a level at which $P_{ST}$ can no longer be sustained, triggering the subsequent relaminarisation.

In order to investigate the temporal relationship between $P_{ST}$ and $\varepsilon_T$, we perform a cross-correlation analysis. For two time-dependent quantities $X(t)$ and $Y(t)$, the cross-correlation is defined as
\begin{equation}
    \hat{\rho}[X,Y](\tau_c)=\frac{\langle X''(t-\tau_c)Y''(t)\rangle}{\sqrt{\langle{X''^2(t)\rangle} ~\langle Y''^2(t)\rangle}},
    \label{eq:xcross}
\end{equation}
where $\langle X\rangle$ denotes a temporal mean and $X''(t)=X(t)-\langle X\rangle$ is the corresponding fluctuation. The timescale $\tau_c$ that maximizes $\hat{\rho}[X,Y]$ provides an average measure of the phase offset between $X$ and $Y$, denoted by $\hat{\tau}$. 

\begin{figure}
  \centering
   \includegraphics[scale =0.28, trim=1.2cm 0.2cm 3.54cm 0cm, clip] {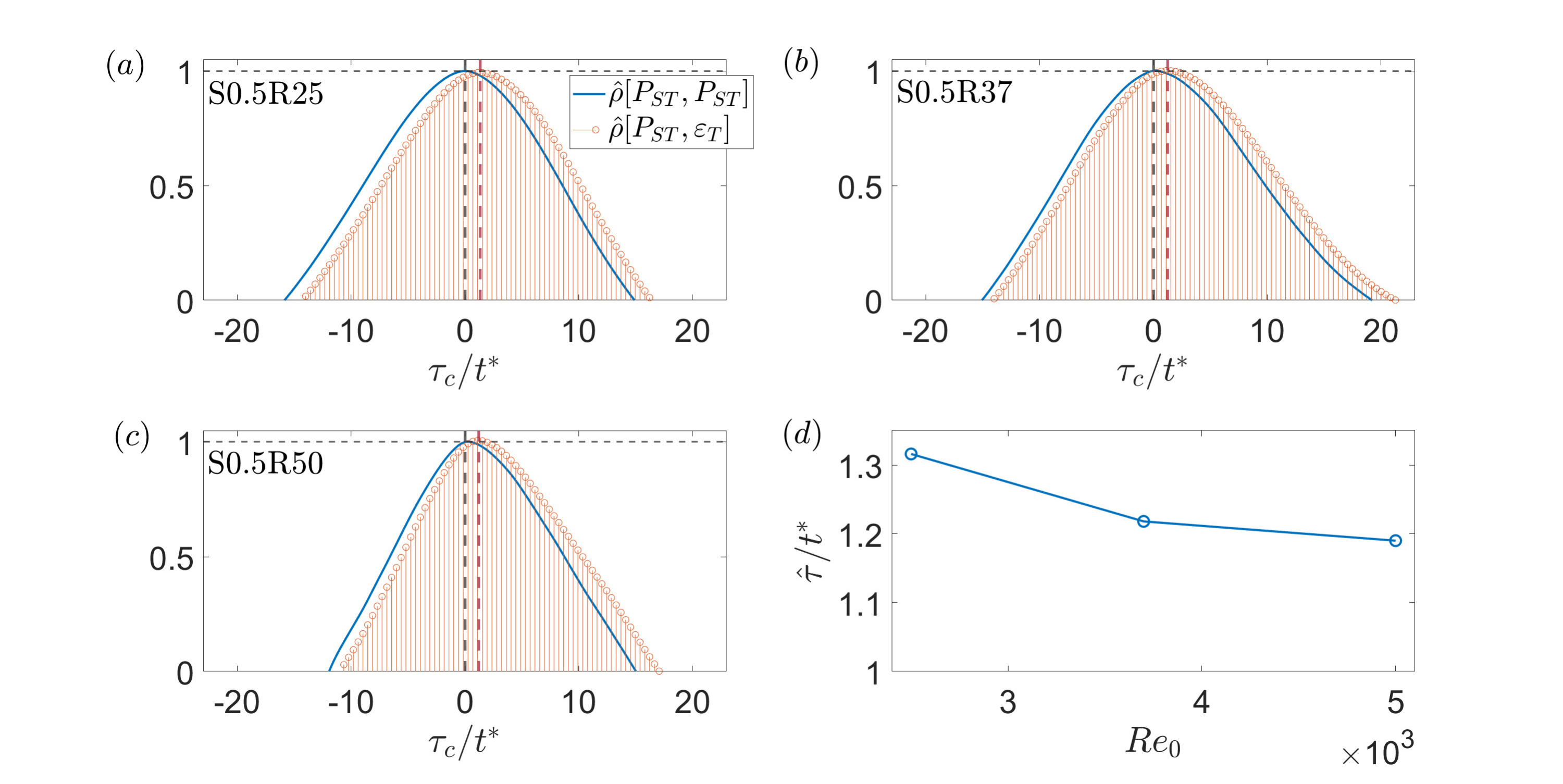}
  \caption{Cross-correlation between $P_{ST}$ and $\varepsilon_T$: $\hat{\rho}[P_{ST}, \varepsilon_T]$, together with the auto-correlation of $P_{ST}$: $\hat{\rho}[P_{ST}, P_{ST}]$ for cases $(a)$ S0.5R25, $(b)$ S0.5R37 and $(c)$ S0.5R50, evaluated over a common correlation window $t/t^*\in(157,230)$. The black vertical lines mark $\tau_c =0$ and the red vertical lines mark the location of $\hat{\tau}/t^*$, which is plotted against $Re_0$ in panel $(d)$. }
 \label{fig:xcorre}
\end{figure}

Figures \ref{fig:xcorre}$(a$--$c)$ present the cross-correlation between production and dissipation, $\hat{\rho}[P_{ST},\varepsilon_T]$ (red hatched areas), for cases S0.5R25, S0.5R37 and S0.5R50, respectively.  A common correlation window of $t/t^*\in(157,230)$ is adopted for all three cases, extending from the restart of the higher-$Re_0$ cases to the end of the first burst event in S0.5R25. The auto-correlation of the production, $\hat{\rho}[P_{ST},P_{ST}]$, is also shown by the blue solid lines, with the vertical black dashed lines marking the location of its maximum at $\tau_c=0$. The cross-correlation $\hat{\rho}[P_{ST},\varepsilon_{T}]$ closely follows the shape of $\hat{\rho}[P_{ST},P_{ST}]$, but is shifted rightward, implying that $\varepsilon_T$ responds to variations in $P_{ST}$ with a finite delay. The red vertical dashed lines mark $\hat{\tau}$ at which $\hat{\rho}[P_{ST}, \varepsilon_T]$ reaches its maximum. Here, $\hat{\tau}$ should be interpreted as a transient response time associated with the burst phase, which dominates the cross-correlation peak. 

The value of $\hat{\tau}$ decreases with increasing $Re_0$, as summarized in figure \ref{fig:xcorre}$(d)$. A shorter delay keeps dissipation more closely coupled to production during transition, potentially limiting transient excursions in TKE and thereby promoting more sustained turbulence at higher $Re_0$. The trend further suggests that $\hat{\tau}$ tends towards a finite asymptotic value as $Re_0$ increases. Such a response time is expected to persist even at high Reynolds numbers, since the small dissipative-scale motions cannot be energised instantaneously as turbulence develops. Additional evidence at higher $Re_0$ will nevertheless be needed to confirm this asymptotic behaviour.

\section{Time-delay system}\label{sec:delaysys}

\subsection{System development}\label{sec:sys}
The dissipation lag and the coupling between the MKE and TKE budgets suggest that the characteristic energetics of the outer layer may be represented within a low-dimensional time-delay dynamical system. We therefore consider a phase space spanned by $\overline{e}_{T}$ and $e_{T}$ and express \eqref{eq:debtdt} and \eqref{eq:detdt} in a compact vector form
\begin{equation}
\frac{\d \mathcal{K}}{\d t}
=-\frac{1}{h}\frac{\d h}{\d t}\mathcal{K}
+ \vec C + \vec B - \vec \varepsilon,
\label{eq:system}
\end{equation}
where 
$\vec{\mathcal{K}}=\begin{bmatrix} \overline{e}_{T} \\ e_{T} \end{bmatrix}$ is the state vector, $\vec C = \begin{bmatrix} -1 \\ 1 \end{bmatrix} P_{ST}$, $\vec B =  \begin{bmatrix} \overline{P}_{BT} \\ P_{BT} \end{bmatrix}$ and $\vec \varepsilon = \begin{bmatrix} 0 \\{\varepsilon}_{T} \end{bmatrix}$.

\begin{figure}
  \centering
   \includegraphics[scale =0.45, trim=11.2cm 0.8cm 11.2cm 13cm, clip] {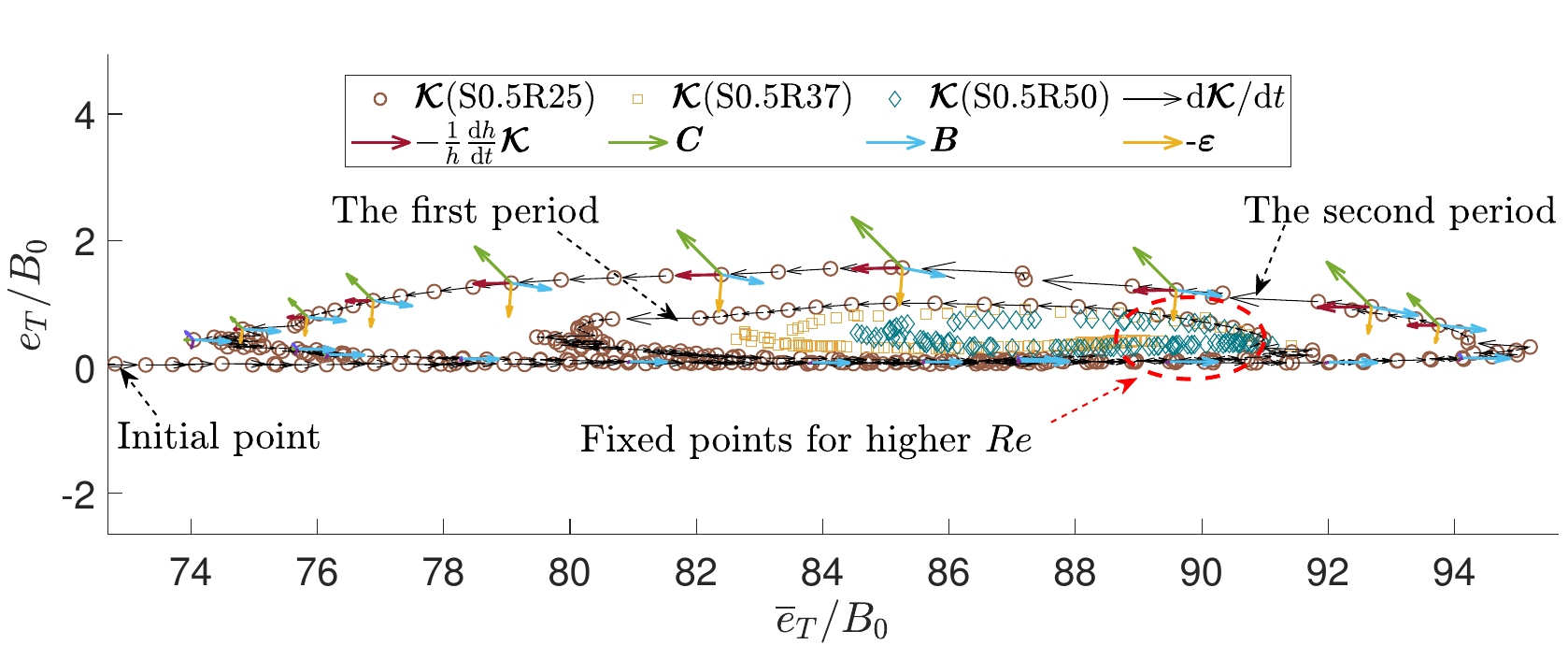}
  \caption{ Phase-space trajectories of the system defined in \eqref{eq:system} for cases S0.5R25, S0.5R37 and S0.5R50. Vectors denote the contributions from individual terms in \eqref{eq:system} governing $\vec{\mathcal{K}}$ for case S0.5R25. The red vectors denoting $-(h^{-1}\d h/\d t)\mathcal{K}$ may partially overlap with the black vectors representing $\d\mathcal{K}/\d t$.}
 \label{fig:phasespace}
\end{figure}

Figure \ref{fig:phasespace} shows phase-space trajectories at different Reynolds numbers. For case S0.5R25 (brown circles), the state vector $\mathcal{K}$ follows orbit-like trajectories, with the second cycle exhibiting stronger oscillations than the first. This behaviour is consistent with the intermittent burst and relaminarisation events  shown in figure \ref{fig:TKEevolution}$(b)$. By contrast, the higher-$Re$ cases S0.5R37 (yellow squares) and S0.5R50 (green diamonds) display progressively weaker oscillations and converge towards fixed points (circled by red dashed lines) representing a turbulent self-similar state, as also evidenced in figures \ref{fig:DifReEvo_2D}$(b)$ and $(c)$. 

The vectors in figure \ref{fig:phasespace} represent the contributions of the individual terms in \eqref{eq:system} to the evolution of $\mathcal{K}$ for case S0.5R25. The trajectory originates from a laminar state, where the system is driven predominantly rightward by the buoyancy production term $\boldsymbol{B}$ that is nearly horizontal since $\overline{P}_{BT}$ dominates. As $\overline{e}_T$ increases, the system transitions towards a turbulent state (corresponding to the upper half of the trajectory), as the conversion term $\boldsymbol{C}$ transfers energy from mean to turbulent kinetic energy by shear production $P_{ST}$. 

In the turbulent state,  $\boldsymbol{C}$ contributes equally and oppositely to $\overline{e}_T$ and $e_T$, pointing diagonally toward the upper left. The dilution term, $-(1/h)(\mathrm{d}h/\mathrm{d}t)\mathcal{K}$, is oriented approximately leftward, as it primarily reduces $\overline{e}_T$. Together, these two terms act to decrease $\overline{e}_T$ and redirect the trajectory back towards lower mean energy states. The dissipation term $-\boldsymbol{\varepsilon}$ points nearly vertically downward, consistent with the negligible dissipation of MKE in the outer layer. The buoyancy production term $\boldsymbol{B}$ develops a downward component in the turbulent state, reflecting the reduction of turbulence through $P_{BT}$. The combined effect of these processes returns the system to a laminar state, after which the cycle repeats under sustained buoyancy forcing.

To close the dynamical system, we parametrise the unresolved production and dissipation terms as
\begin{equation*}
P_{ST}=c_m\frac{e_{T}u_{T}}{h},\quad
\varepsilon_{T}(t) = c_pP_{ST}(t-\tau),\quad
P_{BT}=-\hat{Ri_f}P_{ST},\quad
\refstepcounter{equation}
\eqno{(\theequation{a-c})}
\label{eq:scaleTur}
\end{equation*}
where $c_m$ and $c_p$ are empirical coefficients to be determined, and $\hat{Ri_f}$ denotes a mean outer-layer flux Richardson number. For the small-slope cases considered here, $c_p+\hat{Ri_f}\approx1$ is expected because the dilution term is relatively weak in \eqref{eq:detdt}, so that the energy supplied by $P_{ST}$ is removed primarily through $P_{BT}$ and $\varepsilon_T$. The closure for $P_{ST}$ is based on the outer-layer scalings $ \overline{w'u'} \sim e$ and $\lvert\partial \overline{u}/\partial z\rvert\sim u_T/h$, which have been validated in the self-similar regime by \citet{van2019,cui2025}. We show below that these scalings remain applicable in the intermittent regime. The dissipation $\varepsilon_T$ is modelled as a delayed response to $P_{ST}$, characterising the finite time required for the smallest dissipative scales to adjust to changes in the mean flow. The buoyancy production $P_{BT}$ is represented as a non-delayed fraction of $P_{ST}$ through $\hat{Ri_f}$ \citep{deusebio2015intermittency}. Although irreversible mixing ultimately involves small-scale molecular processes, the DNS data show below that $P_{BT}$ remains closely correlated with $P_{ST}$, suggesting that the buoyancy-flux response is already captured by the shear-production timescale.

\begin{figure}
  \centering
   \includegraphics[scale =0.3, trim=3cm 12.5cm 3cm 0cm, clip] {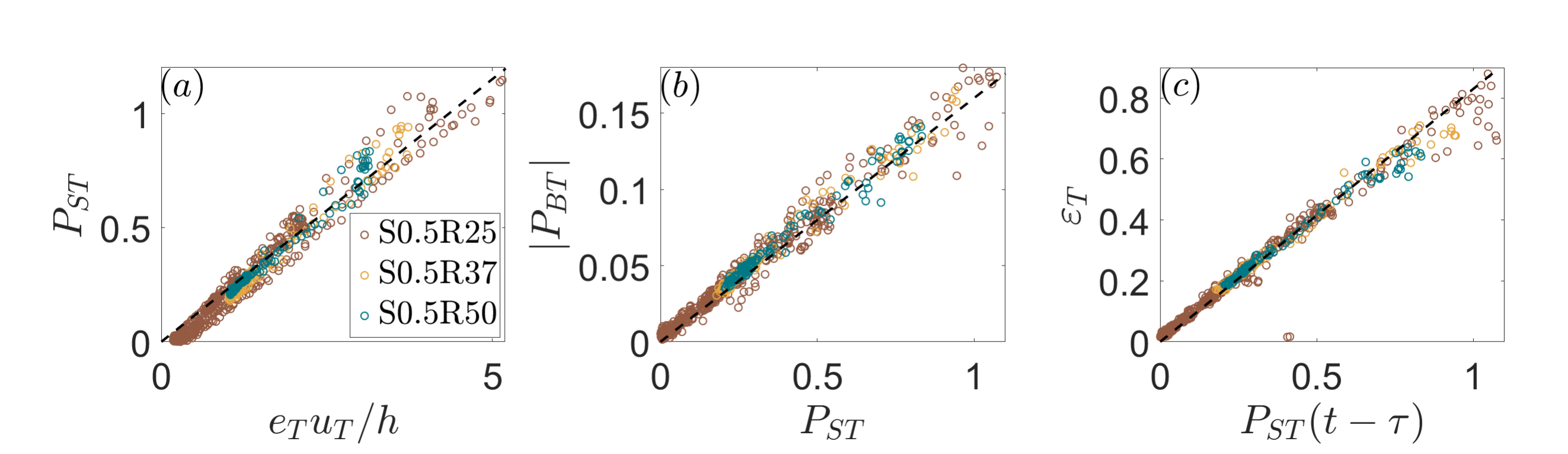}
  \caption{Scaling relations for the outer-layer budget terms: $(a)$  $P_{ST}$, $(b)$  $P_{BT}$ and $(c)$  $\varepsilon_T$ defined in \eqref{eq:scaleTur}. In panel $(c)$, $\tau$ is set to the corresponding value of $\hat{\tau}$ obtained from the cross-correlation analysis for each case. The black dashed lines show the best-fit relations.}
 \label{fig:turscaling}
\end{figure}

Figure~\ref{fig:turscaling} demonstrates the validity of the scaling in \eqref{eq:scaleTur} across the range of Reynolds numbers considered at $\alpha=0.5^\circ$. The delay time $\tau$ in panel $(c)$ is prescribed as the corresponding mean value $\hat{\tau}$ obtained from the cross-correlation analysis. The budget terms collapse onto nearly linear relations, indicating that the parameterisation remains robust within the intermittent regime. The black dashed lines show the corresponding best-fit relations, yielding $c_m \approx 0.23$, $\hat{Ri_f} \approx 0.16$, and $c_p \approx 0.83$. These values agree closely with those reported by \cite{van2019,cui2025} and satisfy $c_p+\hat{Ri_f}\approx1$.  The temporal variations of the predicted $P_{ST}, P_{BT}$ and $\varepsilon_T$ using \eqref{eq:scaleTur} based on the DNS-extracted $e_T$, $u_T$, and $h$ are shown in figure \ref{fig:ebartet} by the solid lines. The predictions show good quantitative agreement with the corresponding DNS budget terms.

Substituting the closure \eqref{eq:scaleTur} into the governing system \eqref{eq:system} yields an autonomous delay system:
\begin{align}
\label{eq:dutdt_final}
  \frac{\d u_{T}}{\d t}&= \frac{B}{h}- \frac{2{c_m}e_{T}}{h},\\
\label{eq:detdt_final}
 \frac{\d e_{T}}{\d t}&= (1-\hat{Ri_f})P_{ST}-c_pP_{ST}(t-\tau)-2{c_m}\frac{e_{T}^2}{u_{T}h},\\
 \label{eq:dhdt_final}
 \frac{\d h}{\d t} &= 2{c_m} \frac{e_{T}}{u_{T}}.
\end{align}
Note that $P_{ST}$ is an explicit function of $u_T, e_T$ and $h$, as given in (\ref{eq:scaleTur}$a$). Equation \eqref{eq:dutdt_final} may be viewed as a reformulation of the MKE budget in velocity form, where the balance between buoyancy input and shear transfer to turbulence appears as a competition between buoyancy forcing and an effective turbulent drag. Equation \eqref{eq:detdt_final} captures the delayed response of dissipation to prior energy input, introducing the finite-time lag between production and dissipation. 

Equation \eqref{eq:dhdt_final} follows from \eqref{eq:dhdt} by substituting the expression for $P_{ST}$. This equation defines the instantaneous layer growth rate without invoking self-similarity, and therefore remains applicable in the intermittent regime. Note that this relation is compatible with the definition of $h$, as can be verified by differentiating $h$ with respect to time and substituting the rate equations for $Q$ and $M$ \citep[implying that this equation is consistent with the entrainment relation; cf.][]{van2015energy}. It further implies that the entrainment rate, $E = u_T^{-1} \d h/\d t = 2c_me_T/u_T^2$, is proportional to the squared turbulence intensity. We note that under self-similar conditions, the ratio $e_T/ u_T^2$ depends on the Richardson number and the classical self-similar behaviour is retrieved. 
\subsection{Parametric study}\label{sec:results}
We now turn to numerical solutions of the delay-differential system to examine its dynamical behaviour. 
A key modelling ingredient is the specification of the delay time $\tau$ in (\ref{eq:scaleTur}$b$). The most direct implementation is to use the measured value $\hat{\tau}$ obtained from the cross-correlation analysis, which represents the effective response time during the turbulent bursts. However, extending this constant delay into the relaminarisation phase can leave dissipation finite after turbulence has collapsed, leading to unphysical negative values of $e_T$. To avoid this unphysical behaviour while retaining $\hat{\tau}$ as the response time during turbulent bursts, we introduce a smooth state-dependent regularisation of $\tau$ based on $\hat{\tau}$:
\begin{equation}
    \tau = \hat{\tau}\tanh\left(c_\tau\frac{e_T}{B}\right),
    \label{eq:taustate}
\end{equation}
where $c_\tau$ is a constant. This form ensures that $\tau$ decreases smoothly as $e_T \rightarrow 0$ and rapidly recovers $\hat{\tau}$ as  $e_T$ begins to increase. Owing to the weak coupling between the inner and outer layers \citep{cui2025}, the buoyancy forcing $B$ is prescribed as a constant, with its value taken from the prediction in that study. Time integration of the system is performed using a fourth-order Runge--Kutta scheme. 

\begin{figure}
  \centering
     \includegraphics[scale =0.3, trim=3cm 2cm 3.5cm 1cm, clip]{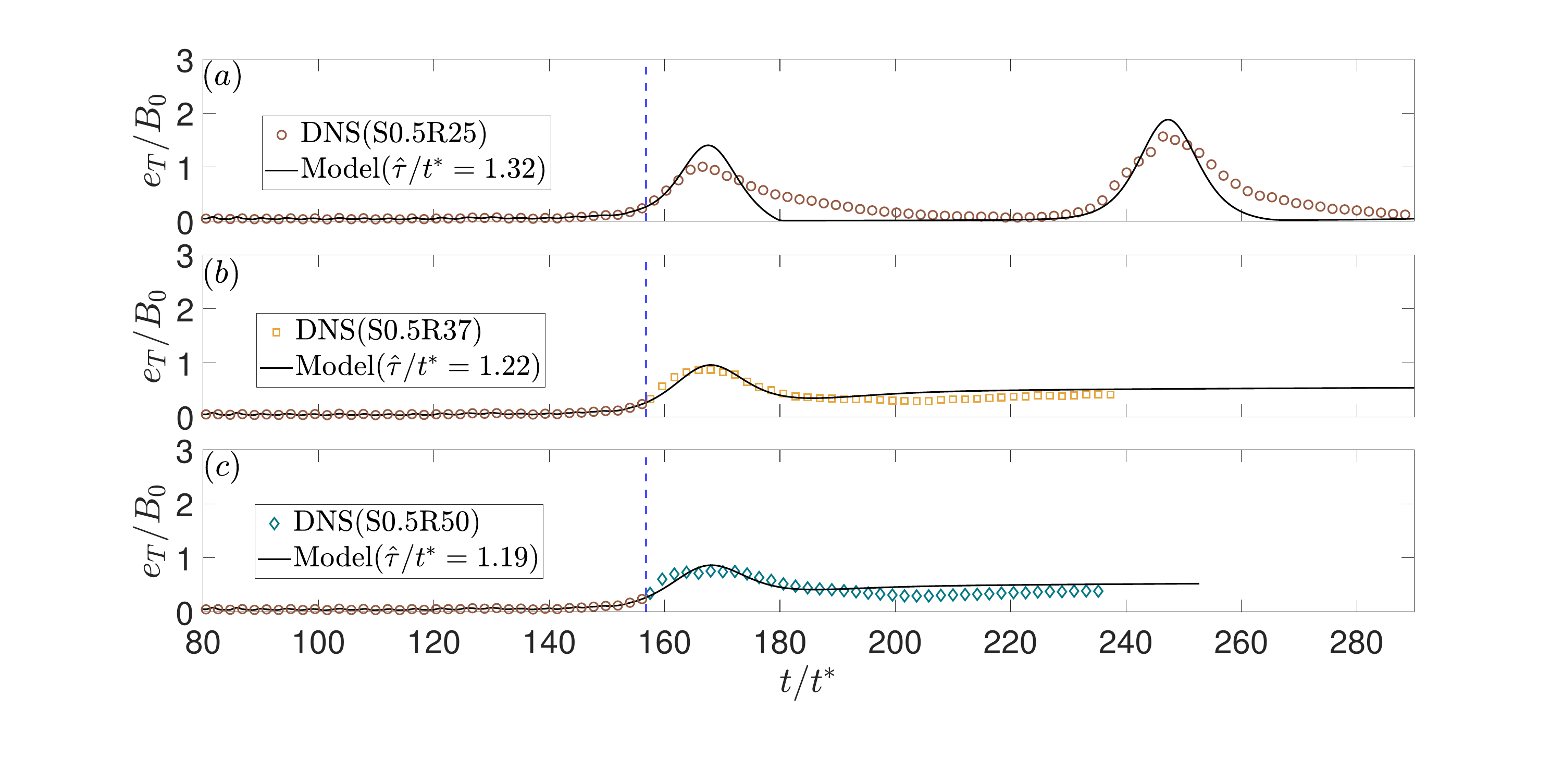}
     \caption{Temporal responses of the proposed time-delay system for varying $\hat{\tau}$, shown together with DNS results for cases $(a)$ S0.5R25, $(b)$ S0.5R37  and $(c)$ S0.5R50. The blue vertical lines mark where we initialise the system.}
     \label{fig:model}
\end{figure}

 Figure \ref{fig:model} shows the temporal variation of the characteristic TKE $e_T$ predicted by the delay-differential system, using $c_\tau=10$ and the values of $\hat{\tau}$ obtained from the cross-correlation analysis at each $Re_0$. The model predictions (solid black lines) are compared with the DNS data at the corresponding $Re_0$ (symbols). The system is initialised at $t/t^* = 157$ (vertical blue lines) where we restart the simulations at higher $Re_0$, and the preceding portion of each curve is extracted from the DNS data. The system exhibits a pronounced sensitivity to the imposed delay. For larger $\hat{\tau}$, the solution develops strong oscillations with multiple peaks, consistent with the turbulent bursts and relaminarisation cycles observed at lower Reynolds numbers. As $\hat{\tau}$ is reduced, these oscillations gradually diminish and the trajectories approach a statistically steady state.

For case S0.5R25 shown in figure \ref{fig:model}$(a)$, the model with $\hat{\tau}/t^* = 1.32$ captures the oscillatory dynamics well, including the timing and magnitude of both burst events. Since the second burst lies outside the correlation window used to estimate $\hat{\tau}$, the agreement between the model and DNS for this event can be viewed as an out-of-sample test of the model. Nevertheless, the system somewhat over-predicts the rate of energy decay as turbulence weakens, likely because the prescription \eqref{eq:taustate} does not fully capture the state dependence of $\tau$. For the higher-$Re_0$ cases shown in panels $(b)$ and $(c)$, $\hat{\tau}/t^* = 1.22$ and $\hat{\tau}/t^* = 1.19$ yield improved agreement with the DNS data at $Re_0=3700$ and $Re_0=5000$, respectively, presumably because the flow is less oscillatory. The agreement between the model predictions and the DNS data further supports the interpretation that increasing $Re_0$ reduces the effective delay associated with dissipation adjustment.

To further explore the parameter dependence of the intermittent dynamics, we rewrite the system \eqref{eq:dutdt_final} -- \eqref{eq:dhdt_final} into a dimensionless form:
 \begin{align}
\label{eq:dutdt_dimless}
  \frac{\d \tilde{u}_{T}}{\d \tilde{t}}&= \frac{1}{\tilde{h}}- 2{c_m}\frac{\tilde{e}_{T}}{\tilde{h}},\\
\label{eq:detdt_dimless}
 \frac{\d \tilde{e}_{T}}{\d \tilde{t}}&= (1-\hat{Ri_f})\tilde{P}_{ST}-c_p\tilde{P}_{ST}(\tilde{t}-\tau/t^*)-2{c_m}\frac{\tilde{e}_{T}^2}{\tilde{u}_{T}\tilde{h}},\\
 \label{eq:dhdt_dimless}
 \frac{\d\tilde{h}}{\d \tilde{t}} &= 2{c_m} \frac{\tilde{e}_{T}}{\tilde{u}_{T}},
\end{align}
where $\tilde{u}_{T} = u_T/\sqrt{B}$, $\tilde{t}=t/t^*$,  $\tilde{h} = h/(t^*\sqrt{B})$, $\tilde{e}_T = e_T/B$, and $\tilde{P}_{ST} = c_m\tilde{e}_T\tilde{u}_{T}/\tilde{h}$. The parameters $c_m$ and $c_p$ may be viewed as fixed constants \citep{van2019, cui2025}. Using the delay closure in \eqref{eq:taustate}, the remaining parameter dependence of the normalised system can be represented in a two-parameter space spanned by the flux Richardson number $\hat{Ri_f}$ and the normalised delay $\hat{\tau}/t^*$.
To quantify intermittency, we define a timescale $T_i$ as 
\begin{equation}
    T_i = T_{m2} - T_{m1},
    \label{eq:T_i}
\end{equation}
where $T_{m1}$ and $T_{m2}$ are the predicted times of the first two local maxima of $e_T$ after the restart point. As illustrated in figure \ref{fig:model}, more pronounced intermittency corresponds to a larger $T_i$, reflecting a longer laminar phase prior to a subsequent burst. When only a single peak is predicted, as for $\hat{\tau}/t^* = 1.19$ in figure \ref{fig:model}, we set $T_i=0$.

\begin{figure}
  \centering
     \includegraphics[scale =0.29, trim=1cm 0cm 1.5cm 0cm, clip] {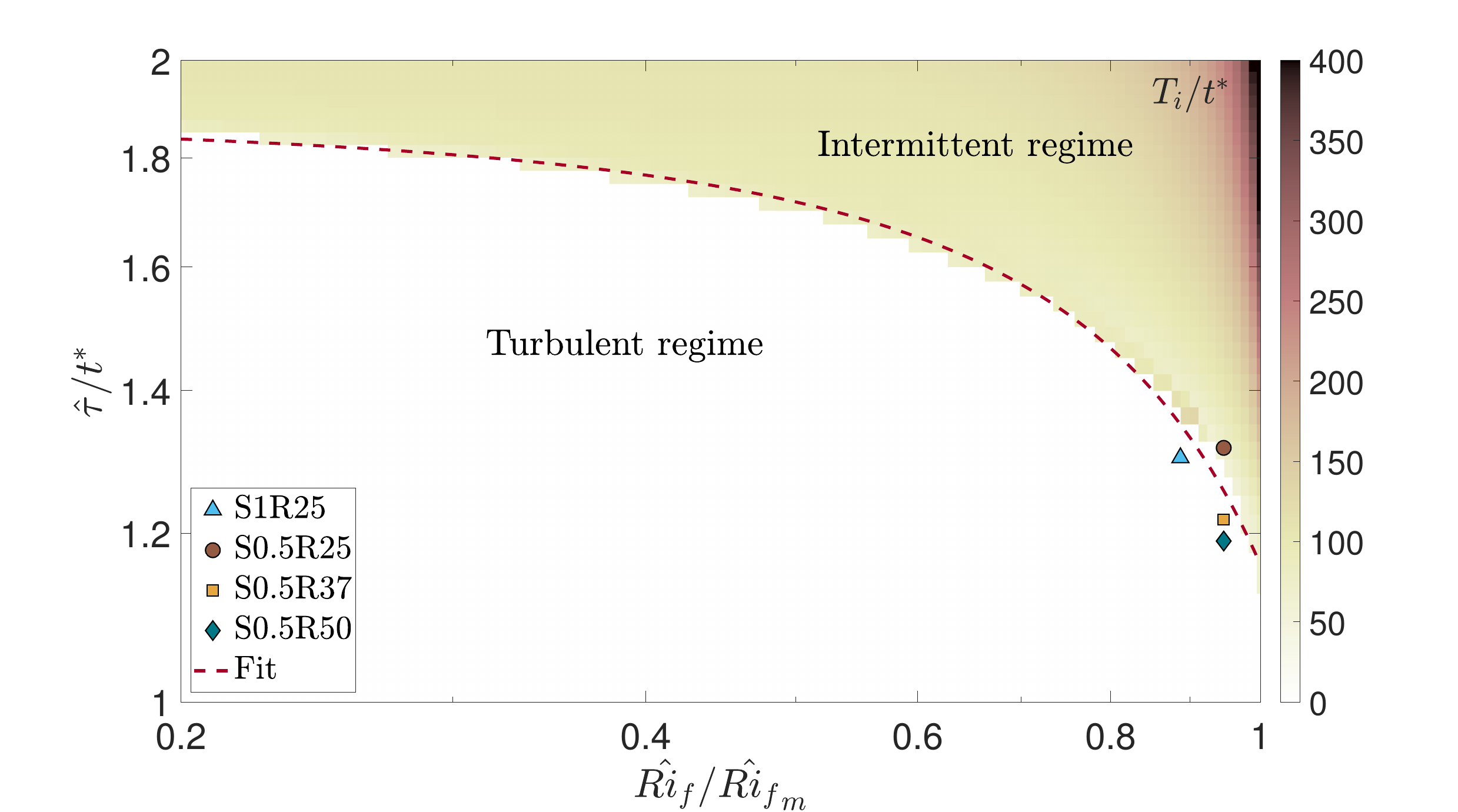}
     \caption{Phase diagram of $T_i/t^*$ as a function of $\hat{Ri_f}/\hat{Ri_f}_m$ and $\hat{\tau}/t^*$. The intermittency boundary (red dashed line) is fitted by $\hat{\tau}/t^* = c_1 \exp\left(c_2 \hat{Ri_f}/\hat{Ri_f}_{m}\right) + c_3$, where $c_1 = -0.07$, $c_2 = 2.45$ and $c_3 = 1.95$. }
     \label{fig:Ti}
\end{figure}

Figure \ref{fig:Ti} shows the phase diagram of $T_i/t^*$ in the $\hat{Ri_f}/\hat{Ri_f}_{m}$--$\hat{\tau}/t^*$ plane predicted by the reduced system, where $\hat{Ri_f}_m = 1-c_p\approx0.17$ denotes the maximum $\hat{Ri_f}$ that permits turbulence to develop in the system. A clear intermittency boundary is seen to separate regimes with $T_i=0$ from those with $T_i>0$, and is well approximated by $\hat{\tau}/t^* = c_1 \exp\left(c_2 \hat{Ri_f}/\hat{Ri_f}_{m}\right) + c_3$, where $c_1 = -0.07$, $c_2 = 2.45$ and $c_3 = 1.95$. The observations from the four DNS cases are indicated by filled symbols. Among the DNS cases, only S0.5R25 lies within the intermittent regime ($T_i > 0$) as expected. 

Longer intermittency times are associated with larger $\hat{Ri_f}$ and $\hat{\tau}/t^*$, indicating that gravity currents become more prone to intermittency at smaller slopes, where $\hat{Ri_f}$ is larger, and at lower Reynolds numbers, where the dissipation lag is larger. This is consistent with the present DNS data and with the results reported by \citet{salinas2020intemigravity,zuniga2022newintermi}. At $\hat{Ri_f}=\hat{Ri}_{fm}$, the critical time delay for the emergence of intermittent turbulence is predicted to remain finite, with $\hat{\tau}\approx1.14t^*$. This finite critical delay supports the possibility that the dissipation lag approaches a non-zero lower bound rather than vanishing as the Reynolds number increases, given that intermittency can persist in submarine gravity currents at very high Reynolds numbers. This interpretation is also consistent with the asymptotic trend observed in figure \ref{fig:xcorre}$(d)$. Establishing the Reynolds-number dependence of this delay more rigorously remains an important direction for future research.

\section{Discussion}
\label{sec:discuss}

The relation $\varepsilon_T(t)\sim P_{ST}(t-\tau)$ reveals a finite delay between turbulent production and dissipation, with the delay decreasing as the initial Reynolds number $Re_0$ increases. This provides a dynamical explanation for the observed intermittency and suggests that the outer layer operates in a fundamentally non-equilibrium regime, where turbulence responds to changes in the mean flow over a finite timescale. At lower Reynolds numbers, this delayed adjustment allows turbulence to amplify transiently, leading to large excursions in turbulence intensity and eventual relaminarisation. As $Re_0$ increases, the adjustment delay decreases and the flow approaches the dynamically equilibrated turbulent regime identified by \citet{cui2025}, in which production and dissipation remain in  balance.

The observed production--dissipation imbalance shares similarities with non-equilibrium turbulent flows, where dissipation does not instantaneously adjust to changes in energy input \citep{vassilicos2015ARFM,valente2012cascade_prl,dairay2015cascade,goto2016cascade}. In the present flow, however, the intermittency is also accompanied by substantial entrainment-driven layer growth, which may affect the magnitudes of the characteristic quantities in a non-trivial way. The measured delay between $P_{ST}$ and $\varepsilon_T$ should therefore be interpreted more broadly as a turbulence-adjustment timescale, characterising the finite response of turbulence to changes in the mean flow. While inter-scale transfer may contribute to this delay, the present simulations do not allow the underlying mechanisms to be isolated.

Intermittency is also associated with variations in the flux Richardson number $\hat{Ri_f}$,
with the most pronounced bursts occurring at the largest values. This is consistent with previous laboratory, numerical and field studies showing that (relatively) strongly stratified gravity currents frequently exhibit intermittent mixing \citep{sun2002intermittent,rorai2014busrt,wells2010relationship}. Although buoyancy production contributes only weakly to the total TKE budget, stratification exerts a disproportionate influence on the dynamics by regulating the shear required to sustain turbulence. This regulation may be linked to internal gravity waves, as discussed by \citet{rorai2014busrt}. Consistent with this, wave-like signatures are observed during the relaminarisation phase following the initial turbulent burst (figure~\ref{fig:enstrophy} $(d)$).

The intermittency observed here is closely related to the accelerating and decelerating regimes identified by \citet{salinas2020intemigravity} and \citet{zuniga2022newintermi}. The present results suggest that these regimes correspond to different phases of a relaxation cycle. Buoyancy forcing accelerates the current and increases the mean shear during quiescent phases; once turbulence is triggered, enhanced momentum transfer and entrainment weaken the mean flow and shear until turbulence can no longer be sustained, leading to relaminarisation and, in turn, to  renewed acceleration.

Intermittent turbulence in these stratified gravity currents also shares conceptual similarities with subcritical transitions and non-equilibrium statistical phase transitions \citep{wygnanski1973puff-slug1,wygnanski1975puff-slug2,mullin2011transition,avila2023transition}. While a classical turbulent--laminar front is not well defined in the present temporal configuration, the alternation between active and quiescent phases is qualitatively reminiscent of directed-percolation dynamics \citep{pomeau1986DPorigin}, where the persistence of activity depends on the competition between proliferation and decay. Establishing whether this connection extends beyond a qualitative analogy would require substantially larger domains and a detailed analysis of the spatial organisation of intermittency.

Taken together, the present results suggest that the intermittency observed by \citet{salinas2020intemigravity,zuniga2022newintermi} and the dynamically equilibrated regime identified by \citet{cui2025} are not distinct flow states but different dynamical responses of the same system. At sufficiently large Reynolds numbers, turbulence adjusts rapidly enough for production and dissipation to remain approximately balanced. At lower Reynolds numbers, the finite turbulence-adjustment time leads to repeated turbulent burst and relaminarisation events, producing the observed intermittent behaviour. Viewed in this way, the intermittency may be interpreted as a delay-induced relaxation oscillation arising from the finite response time of turbulence to changes in the mean flow.

\section{Conclusion}\label{sec:conclusion}
In this study, we have investigated the physical origin of intermittent turbulence in inclined gravity currents at shallow slopes using direct numerical simulations over a range of initial Reynolds numbers $Re_0$. At a slope of $\alpha=0.5^\circ$, the outer-layer dynamics display a clear sensitivity to $Re_0$: for the lowest value, the flow does not directly settle into the dynamically equilibrated regime reported by \cite{cui2025}, but instead exhibits pronounced intermittency. As $Re_0$ increases, the intermittency progressively weakens and the flow approaches a more sustained turbulent state. By contrast, the inner layer remains fully turbulent in all cases and shows no appreciable dependence on $Re_0$. Our analysis therefore focusses on the outer layer, where the intermittent dynamics emerge.

The outer-layer intermittency consists of recurrent transitions between turbulent bursts and relaminarisation phases. During quiescent periods, buoyancy forcing increases the mean kinetic energy (MKE) and strengthens the mean shear. Once turbulence is triggered, MKE is transferred to turbulent kinetic energy (TKE) through shear production, forming the primary energetic coupling between the two reservoirs. This transfer is accompanied by a geometric dilution term associated with entrainment-driven layer growth, and these two contributions to MKE are found to be approximately equal. Hence turbulence rapidly and significantly weakens the shear on which it relies. In the TKE budget, the leading-order balance is between shear production and dissipation; however, this is not an instantaneous equilibrium. Dissipation responds to variations in production with a finite lag, producing temporary imbalances during transitional phases that can lead to transient amplification of TKE. This amplified turbulence subsequently weakens the mean shear, suppresses production, and returns the flow to a more quiescent or weakly turbulent state.

The magnitude of this lag decreases systematically with increasing $Re_0$, as indicated by the cross-correlation analysis. This behaviour explains why intermittency weakens at higher Reynolds numbers: shorter delays diminish the transient mismatch between production and dissipation, allowing the flow to transition more directly into sustained turbulence. While the detailed physical origin of the delay remains unclear, the present results demonstrate that its magnitude plays a central role in controlling the transition between intermittent and dynamically equilibrated turbulence.

By incorporating this delayed dissipation response into the outer-layer energetics, we developed a low-dimensional delay-differential system with MKE and TKE as the state variables. When the prescribed delay is varied, the system reproduces the DNS transition from strongly oscillatory behaviour at low $Re_0$ to weakly oscillatory or sustained turbulence at higher $Re_0$. The delay system also 
suggests that larger $\hat{Ri_f}$, corresponding to stronger stratification, is associated with intermittency and stronger burst events. The DNS further indicates that stronger stratification allows larger amounts of MKE to accumulate during quiescent phases. Once turbulence develops, this energy is rapidly transferred to turbulence and, in combination with the delayed dissipation response, produces more intense burst events.

The present results support an interpretation of the outer-layer intermittent turbulence in inclined gravity currents as a delay-induced oscillation. Several questions, however, remain open. At relatively small slope angles (larger $\hat{Ri_f}$), the long-time behaviour is uncertain: it is not yet clear whether the outer layer may remain permanently intermittent,  settle into a laminar regime as suggested by \cite{salinas2021nature}, or eventually transition into the dynamically equilibrated regime reported by \cite{cui2025}. In addition, the origin of the production--dissipation delay and its dependence on Reynolds number and stratification warrant further investigation beyond the simplified closure adopted here. It also remains to be tested whether the same delay-controlled mechanism persists in spatially developing gravity currents and more realistic geophysical settings.

\section*{Funding}
The authors acknowledge the UK Turbulence Consortium (EPSRC grants EP/R029326/1 and EP/X035484/1) for the grand challenge project that provided the computational resources for this work. Lianzheng Cui acknowledges the Skempton Scholarship and financial support from the China Scholarship Council for his PhD study. Graham O. Hughes acknowledges support from the Engineering and Physical Sciences Research Council (grant EP/V033883/1).

\section*{Declaration of interests}
The authors report no conflict of interest.

\section*{Data availability statement}
The data that support the findings will be made openly available upon publication.

\appendix
\section{Characteristic buoyancy production of mean kinetic energy}
\label{sec:barPBT}
We assume that the mean streamwise velocity $\overline{u}$ and mean buoyancy $\overline{b}$ are coupled and share a common vertical structure when normalised by their respective characteristic scales. Specifically, we write
\begin{equation}
\overline{u} = u_T f_{ub}(\eta),
\qquad
\overline{b} = b_T f_{ub}(\eta),
\end{equation}
where $f_{ub}$ is a dimensionless profile and $\eta = (z - z_{um})/h$. The buoyancy production of mean kinetic energy (MKE) $\overline{P}_{B}$ is given by
\begin{equation}
    \overline{P}_{B} =-\overline{u}\overline{b}\sin\alpha = -u_{T}f_{ub}(\eta)b_{T}\sin\alpha f_{ub}(\eta) = \frac{Bu_{T}}{h} f_{ub}^2(\eta).
\end{equation}
The characteristic buoyancy production of MKE $\overline{P}_{BT}$ is then defined as
\begin{equation}
    \overline{P}_{BT} =\frac{1}{h}\int_{z_{um}}^\infty \overline{P}_{B} \d z= \frac{Bu_{T}}{h}\int_0^{\eta_1} f_{ub}^2(\eta) \d \eta,
\end{equation}
where $\eta_1$ denotes the zero-crossing location of the  function $f_{ub}$. Recalling the definition of the momentum flux $M$, we have
\begin{equation}
    M=\int_{z_{um}}^\infty \overline{u}^2 \d z =u_T^2h\int_0^{\eta_1} f_{ub}^2(\eta) \d \eta= u_T^2h\quad\Rightarrow\quad \int_0^{\eta_1} f_{ub}^2(\eta) \d \eta=1.
\end{equation}
Substituting into the expression for $\overline{P}_{BT}$ yields
\begin{equation}
\overline{P}_{BT} = {B u_T}/{h}.
\end{equation}
 
\bibliographystyle{plainnat}
\bibliography{references}

\end{document}